\documentclass[journal]{IEEEtran}
\usepackage[cmex10]{amsmath}
\usepackage{amsfonts}
\usepackage{amssymb}
\usepackage{algorithmic}
\usepackage{algorithm}
\usepackage{array}
\usepackage{bm}

\usepackage{textcomp}
\usepackage{stfloats}
\usepackage{url}
\usepackage{verbatim}
\usepackage{graphicx}
\usepackage{cite}

\usepackage{soul}
\usepackage[dvipsnames]{xcolor}
\sethlcolor{CornflowerBlue}

\usepackage{multirow}

\usepackage[tight,footnotesize]{subfigure}
\usepackage{subfiles}

\newtheorem{lemma}{Lemma}

\newtheorem{proposition}{Proposition}
\newtheorem{remark}{Remark}

\begin{document}

\title{SIC-Free Rate-Splitting Multiple Access: Constellation-Constrained Optimization and Application to Large-Scale Systems}


\author{
    \IEEEauthorblockN{
        Sibo~Zhang,~\IEEEmembership{Graduate Student Member,~IEEE}, ~Bruno~Clerckx,~\IEEEmembership{Fellow,~IEEE},~David~Vargas
    }
    \thanks{This work was supported in part by the UK Engineering and Physical Sciences Research Council, Industrial Case award number 210163.}
    \thanks{This work was presented in part at 25th IEEE International Workshop on Signal Processing Advances in Wireless Communications, Lucca, Italy, Sep. 2024 \cite{Sibo_SPAWC}.}
    \thanks{S. Zhang is with the Department of Electrical and Electronic Engineering at Imperial College London, London SW7 2AZ, U.K. and BBC Research and Development, The Lighthouse, White City Place, 201 Wood Lane, London, W12 7TQ, U.K. (e-mail: sibo.zhang19@imperial.ac.uk).}
    \thanks{B. Clerckx is with the Department of Electrical and Electronic Engineering at Imperial College London, London SW7 2AZ, U.K. (e-mail: b.clerckx@imperial.ac.uk).}
    \thanks{D. Vargas is with BBC Research and Development, The Lighthouse, White City Place, 201 Wood Lane, London, W12 7TQ, U.K. (e-mail: david.vargas@bbc.co.uk).} 
    }



\maketitle

\begin{abstract}
Rate-Splitting Multiple Access (RSMA) has been recognized as a promising multiple access technique for future wireless communication systems. Recent research demonstrates that RSMA can maintain its superiority without relying on Successive Interference Cancellation (SIC) receivers. In practical systems, SIC-free receivers are more attractive than SIC receivers because of their low complexity and latency. This paper evaluates the theoretical limits of RSMA with and without SIC receivers under finite constellations. We first derive the constellation-constrained rate expressions for RSMA. We then design algorithms based on projected subgradient ascent to optimize the precoders and maximize the weighted sum-rate or max-min fairness among users. To apply the proposed optimization algorithms to large-scale systems, one challenge lies in the exponentially increasing computational complexity brought about by the constellation-constrained rate expressions. In light of this, we propose methods to avoid such computational burden. Numerical results show that, under optimized precoders, SIC-free RSMA leads to minor losses in both weighted sum-rate and max-min fairness in comparison to RSMA with SIC receivers, making it a viable option for future implementations.
\end{abstract}

\begin{IEEEkeywords}
Rate-splitting multiple access (RSMA), multi-antenna broadcast
channel, finite-alphabet signalling, massive MIMO.
\end{IEEEkeywords}

\section{Introduction}
Multiple access techniques have played a pivotal role in the advancement of wireless communication systems. Recent research suggests that Rate-Splitting Multiple Access (RSMA) has significant potential to improve future wireless systems, such as 6G and beyond, due to its numerous benefits, including efficiency, universality, flexibility, robustness, reliability, and low latency \cite{Mao,Clerckx}. The exploration of RSMA is grounded by early work on Rate-Splitting (RS) for two-user single-input-single-output interference channels \cite{Carleial,Han}, and aims at designing advanced multi-user transmission strategies for multi-antenna broadcast channels (downlink) and multiple access channels (uplink). The application of the concept of RS to Multi-User Multiple-Input-Multiple-Output (MU-MIMO) systems, later referred to as RSMA \cite{Mao_EURASIP}, was initiated by \cite{Hamdi2016} to address the problem of partial Channel State Information at the Transmitter (CSIT) in downlink MIMO systems. Since then, research has shown that RSMA benefits a wide range of communication scenarios. For downlink systems, these include, but are not limited to, massive MIMO \cite{Minbo2016MassiveMIMO,Onur_mobility}, integrated sensing and communications \cite{Chengcheng2021isac}, satellite communications \cite{longfei2021satellite}, short-packet systems \cite{yunnuo2023FBL}, joint unicast and multicast \cite{mao2019TCOM}, overloaded networks\cite{onur_MMF} and reconfigurable intelligent surfaces \cite{Hongyu2024}. \cite{Rimoldi} provides a basis for RSMA for single-antenna multiple access channels, or uplink systems. For MIMO uplink systems, the benefits has been exploited in short-packet systems \cite{Jiawei2023} and network slicing \cite{Yuanwen2024}.

Most existing work assumes that a Successive Interference Cancellation (SIC) receiver is essential for RSMA, which is referred to as SIC-based RSMA in this work. However, more recent work in \cite{Sibo2024} initiates research on RSMA without SIC, i.e., SIC-free RSMA\footnote{\cite{Sibo_SPAWC} and \cite{Sibo2024} use the terms "SIC-free" and "non-SIC" interchangeably, as they convey the same meaning. In this work, we adopt the term "SIC-free" for consistency, and refer "RSMA with SIC receivers" as "SIC-based RSMA" for clarity.}, and focuses on its theoretical metric, receiver implementations, and the corresponding precoder designs. The findings in \cite{Sibo2024} suggest that the exclusion of SIC from RSMA receivers reduces complexity, buffer size, and delay and leads to a minor impact on performance. The RSMA implementations discussed in \cite{Sibo2024} are further verified under the physical layer procedures of the 5G New Radio (NR) and the more practical channel models generated from ray-tracing analysis to support the first industrial contribution to RSMA in 3GPP \cite{BBC}. Notably, the precoder designs proposed in \cite{Sibo2024} adopt a low-complexity approach, making it advantageous for practical systems due to its minimal computational burden. Nonetheless, this approach may not fully capture the theoretical limits of SIC-based and SIC-free RSMA. Meanwhile, although RSMA has been evaluated under finite constellations in many works, such as in \cite{Sibo2024, Onur_mobility, Onur_LLS}, optimization of RSMA has rarely been carried out utilizing knowledge of finite constellations. This indicates a need to understand the constellation-constrained theoretical limits of RSMA.

Information theory suggests that Gaussian-distributed input signal is capacity-achieving for a wide range of communication scenarios. However, it is difficult to apply Gaussian signals in practice because of mainly two reasons: 1) their unbounded amplitude imposes hardware challenges; and 2) their continuous support leads to difficulty in coding and detection. As a result, practical systems often apply input signals with finite alphabets, leading to modulation schemes such as Pulse Amplitude Modulation (PAM) and Quadrature Amplitude Modulation (QAM). It is therefore interesting to investigate the design and performance of communication systems under finite constellations, as they are generally different from those for Gaussian distributed input signals. For example, \cite{Lozano} considers parallel Gaussian channels with arbitrary input distributions and proposes a mercury water-filling policy for power allocation as an extension to the classical water-filling policy for Gaussian inputs. The precoder design for point-to-point MIMO systems is considered in \cite{Palomar,Perez-Cruz}. For uplink multi-user systems with finite constellations, \cite{Harshan} studied two-user multiple access channels with finite constellations, while linear precoding for MIMO multiple access channels was addressed in \cite{Girnyk, Wu_MAC_channel, Pei_MAC_channel}. For downlink, \cite{Wu_BC_channel} investigated the precoder design for MIMO broadcast channels. Finite constellations are also considered in more complicated systems, for example, multi-cell systems \cite{Wu_multicell}, cognitive radio \cite{Zeng_cognitive}, energy harvesting \cite{Gregori} and relay networks \cite{Zeng_relay}. Although the above-mentioned works focus mainly on mutual information, designs following different criteria, such as mean squared error and diversity, exist and are surveyed in \cite{Wu_survey}. Note that the computational complexity of mutual information typically increases exponentially with the number of signal sources and the cardinality of constellations in use, causing designs for large-scale systems (e.g., massive MIMO) to be difficult and in need of simplification. For example, \cite{Girnyk} evaluates the asymptotic rate expression when the number of interference sources tends to be very large. \cite{Wu_multicell} uses Gaussian distribution to approximate the sum of finite alphabet signals that interfere. 

In this work, we discover the fundamental limits of SIC-based RSMA and SIC-free RSMA under finite constellations. The key contributions of this work are listed as follows:
\begin{itemize}
    \item We derive the constellation-constrained rate expressions and their approximation for SIC-based and SIC-free RSMA. Note that the RSMA rate expressions derived in \cite{Sibo2024} are based on the assumption that zero-forcing precoders are used for private streams. As a result of this assumption, inter-user interference was removed during the derivations and the rate expressions were simplified. The rate expressions derived in this work do not make any assumption on the precoders and are therefore more general than those in \cite{Sibo2024}.
    \item We propose precoder optimization methods to maximize the Weighted Sum-Rate (WSR) and Max-Min Fairness (MMF) (or minimum rate per user) performance for SIC-based and SIC-free RSMA under finite constellation constraints. To enable this, we developed global optimal solutions to common stream rate allocation for both WSR maximization and MMF problems. These solutions hold for not only finite-constellation inputs but also any other input signal distribution, including Gaussian inputs. 
    \item We observe that the computational complexity of the proposed algorithms grows exponentially as the system scales up in terms of the number of users. In light of this, we propose methods to scale up optimization algorithms to handle large-scale systems and avoid prohibitive computational cost. 
    \item We show through simulations that both SIC-based and SIC-free RSMA outperform SDMA under different evaluation metrics and dimensions of the system. Moreover, SIC-free RSMA often preserves most of the superiority of SIC-based RSMA. The results also show that SIC-based and SIC-free RSMA require very different precoders, despite the fact that they achieve similar performance in the end. This indicates the need for dedicated precoder designs for SIC-free RSMA. 
    Therefore, we extend the conclusion in \cite{Sibo2024} from low-complexity precoders to fully optimized precoders, from the sum-rate objective to the WSR and MMF objectives, and from small-scale to large-scale systems.
\end{itemize}

\emph{Organization:} The rest of this paper is organized as follows. Section II introduces a system model for RSMA with explanations of SIC and SIC-free receivers. In Section III, derivation of achievable rate expressions for RSMA under finite constellations is provided. In Section IV, precoder optimization algorithms for WSR and MMF optimization are proposed. Section V addresses optimization in large-scale systems. Section VI presents and analyzes numerical results. Finally, Section VII concludes this paper.

\emph{Notations:}  Italic, bold lower-case, bold upper-case and calligraphic letters denote respectively scalars, vectors, matrices and multisets. $\mathbf{I}$, $\mathbf{0}$ and $\mathbf{1}$ denote the identity matrix, the zero matrix, the matrix of ones, with dimensionality given by their superscripts. $[\cdot]_{m,n}$ denotes the $(m,n)$-th entry in a matrix. $(\cdot)^T$, $(\cdot)^H$, $\text{tr}(\cdot)$ and $\|\cdot\|$ denote, respectively, the transpose, conjugate transpose, trace, and Euclidean norm of the input entity. $|\cdot|$ denotes the absolute value if the argument is a scalar, or the cardinality if the argument is a set. $\times$ denotes the Cartesian product of two sets. $\mathbb{E}\{\cdot\}$ denotes the expectation. $\mathcal{CN}(\boldsymbol{\mu},\mathbf{\Sigma})$ denotes multivariate circular symmetric complex Gaussian distribution with mean vector and covariance of $\boldsymbol{\mu}$ and $\mathbf{\Sigma}$. Sum and product of multisets are defined using Minkowski sum and Cartesian product, respectively.

\section{RSMA System Model}\label{Section_system_model}
Consider a downlink scenario in which a transmitter (base station) equipped with $N_\text{T}$ antennas serves $K$ single-antenna receivers (users) with unicast messages. Let $\mathcal{K} = \{1,2,...,K\}$ denote the set of user indices. With the assumption on narrow band, the received signal at the $k$-th user can be expressed as
\begin{equation}
    y_k = \mathbf{h}_k^H \mathbf{x} + n_k ,
\end{equation}
where $\mathbf{h}_k \in \mathbb{C}^{N_\text{T} \times 1}$ denotes the channel response vector between the transmitter and the $k$-th user, $\mathbf{x} \in \mathbb{C}^{N_\text{T} \times 1}$ denotes the transmitted signal vector and $n_k \sim \mathcal{CN}(0,\sigma^2)$ denotes the additive white Gaussian noise observed by the $k$-th user. 

\begin{figure*}[h!]
    \centering
    \includegraphics[width=18cm]{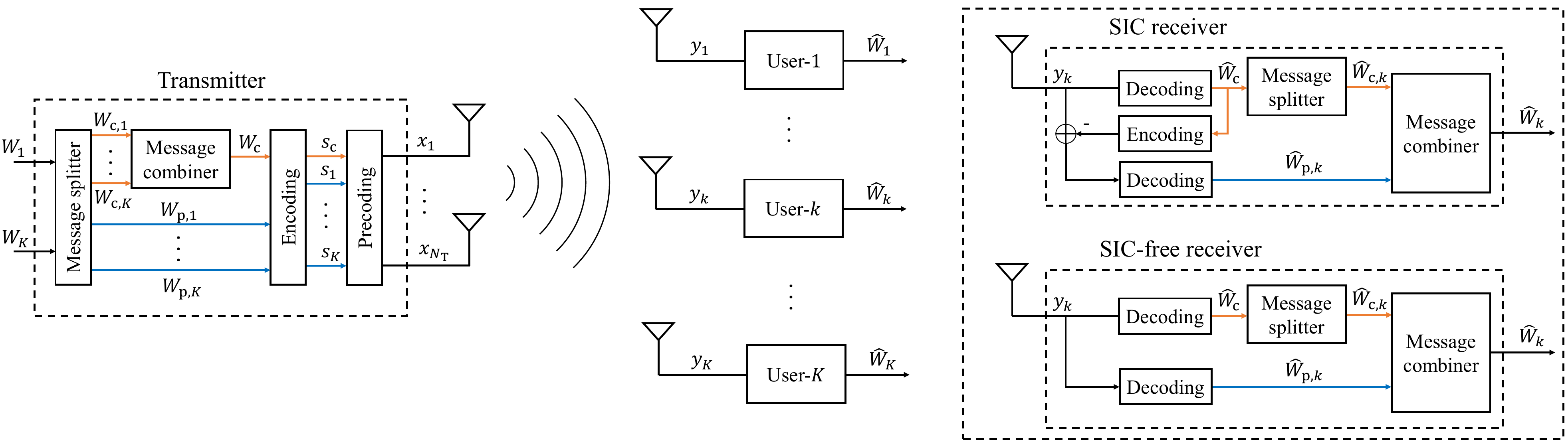}
    \caption{Single-layer RSMA with SIC and SIC-free receivers.}
    \label{RSMA system model}
\end{figure*}

The transmitter applies single-layer RSMA \cite{Mao_EURASIP} as the multiple access scheme, whose key procedures are shown in Figure \ref{RSMA system model} and are explained as follows. Let $\{W_1,...,W_K\}$ respectively denote the $K$ unicast messages to be transmitted to the $K$ users. The transmitter first processes the messages by splitting each message into common and private parts, e.g., $W_k$, $\forall k\in\mathcal{K}$, is split into $W_{\text{c},k}$ and $W_{\text{p},k}$. Next, the $K$ common parts, i.e., $\{W_{\text{c},1},...,W_{\text{c},K}\}$, are combined into one common message, denoted by $W_{\text{c}}$, while the private parts, i.e., $\{W_{\text{p},1},...,W_{\text{p},K}\}$, are treated as individual messages. The resulting $K+1$ messages, $\{W_{\text{c}},W_{\text{p},1},...,W_{\text{p},K}\}$, are then encoded in $K+1$ common and private symbol streams, denoted by $\{s_{\text{c}},s_{1},...,s_{K}\}$. Finally, these symbols are precoded and transmitted. Therefore, we express the transmitted signal as
\begin{equation}
    \mathbf{x} = \mathbf{Ps},
\end{equation}
where $\mathbf{P} = [\mathbf{p}_\text{c}, \mathbf{p}_1, ..., \mathbf{p}_K]$ denotes the precoding matrix for the common and private streams, and $\mathbf{s}=[s_\text{c}, s_1, ..., s_K]^T$. We assume that the symbols are drawn independently and uniformly from finite constellations with zero mean and unit average power. Therefore, $\mathbb{E}\{\mathbf{s}\mathbf{s}^H\} = \mathbf{I}^{K+1}$. 

As for receiver implementation at user-$k$, $\forall k\in\mathcal{K}$, we consider two potential architectures as depicted in Figure \ref{RSMA system model}:
\begin{itemize}
    \item SIC receiver: The receiver first decodes $W_\text{c}$ from $y_k$ by treating $\{s_1,...,s_K\}$ as noise and obtains $\Hat{W}_\text{c}$, denoting the resulting estimate of $W_\text{c}$. $\Hat{W}_\text{c}$ is then re-encoded into $\Hat{s}_\text{c}$, an estimate of the common stream symbol. Next, the receiver decodes $W_{\text{p},k}$ from $y_k - \mathbf{h}_k^H\mathbf{p}_\text{c}\Hat{s}_\text{c}$, i.e., the received signal with the estimated common stream symbol canceled, by treating $s_{k'}$, $\forall k'\in\mathcal{K}/k$, as noise and obtains $\Hat{W}_{\text{p},k}$, an estimate of the desired private message. Finally, the receiver extracts $\Hat{W}_{\text{c},k}$ from $\Hat{W}_\text{c}$ and combines it with $\Hat{W}_{\text{p},k}$ to reconstruct $\Hat{W}_k$ as the resulting estimate of the original unicast message for user-$k$.
    
    Throughout this paper, the following three assumptions are made to facilitate the analysis: \begin{enumerate}
        \item Channel State Information (CSI) is perfectly known at the transmitter and the receivers;
        \item the transmission block is of infinite length;
        \item all the messages are encoded at rates that are decodable at the corresponding receivers.
    \end{enumerate}
    Hence, $\Hat{W}_\text{c} = W_\text{c}$, i.e., the cancellation of common stream is always perfect.

    \item SIC-free receiver: The receiver decodes $W_\text{c}$ from $y_k$ by treating $\{s_1,...,s_K\}$ as noise, and also decodes $W_{\text{p},k}$ by treating $s_\text{c}$ and $s_k'$, $\forall k'\in\mathcal{K}/k$, as noise. After this, the receiver extracts $\Hat{W}_{\text{c},k}$ from $\Hat{W}_\text{c}$ and combines it with $\Hat{W}_{\text{p},k}$ to reconstruct $\Hat{W}_k$. The main difference from the SIC implementation is that $\Hat{s}_\text{c}$ remains in the received signal when the receiver decodes $W_{\text{p},k}$ due to the absence of the cancellation step.
\end{itemize}

Note that Spatial Domain Multiple Access (SDMA), which has been widely used for downlink MU-MIMO, is a special case of RSMA, as RSMA becomes SDMA when $\|\mathbf{p}_\text{c}\|^2=0$. 

\section{Achievable Rate of RSMA Under Finite Constellations}
In this section, we present the mutual information of a linear system with finite alphabet input and utilize it to derive the rate expressions of RSMA.
\subsection{Prerequisites on Constellation-Constrained Rate}
Consider a linear system represented by\footnotemark
\begin{equation}
    y = \mathbf{h}^H\mathbf{P} \mathbf{s} + n.
\end{equation}
\footnotetext{The output is assumed to be a scalar because we focus on MISO systems in the paper. Extension to systems with vector output, such as MIMO systems, is straightforward.}where $n \sim \mathcal{CN}(0,\sigma^2)$, $\mathbf{h}^H\mathbf{P}$ is fixed and known, and $\mathbf{s}$ is an observation of a uniformly distributed random variable with finite support given by $\mathcal{X}$. Let $\mathbf{s}_m$ denote the $m$-th element in $\mathcal{X}$ in a certain order. The mutual information in bits between $\mathbf{s}$ and $y$ is given by
\begin{equation}   I(\mathbf{s};y|\mathbf{h}^H\mathbf{P}) = \log_2(|\mathcal{X}|) - H(\mathbf{s}|y,\mathbf{h}^H\mathbf{P}),
\end{equation}
where $H(\mathbf{s}|y,\mathbf{h}^H\mathbf{P})$ is the conditional entropy of $\mathbf{s}$ given $y$ and $\mathbf{h}^H\mathbf{P}$, and is expressed as \cite{Wu_BC_channel}
\begin{equation}\label{con_entropy}
\begin{split}
    &H(\mathbf{s}|y,\mathbf{h}^H\mathbf{P})\\
    &= \frac{1}{\ln2} + \frac{1}{|\mathcal{X}|} \sum_{m=1}^{|\mathcal{X}|} \mathbb{E}_n \Big\{ \log_2 \sum_{l=1}^{|\mathcal{X}|} \exp \Big( -\frac{|\mathbf{h}^H\mathbf{P} \mathbf{d}_{m,l}(\mathcal{X}) + n|^2}{\sigma^2} \Big) \Big\},
\end{split}
\end{equation}
where $\mathbf{d}_{m,l}(\mathcal{X}) = \mathbf{s}_m - \mathbf{s}_l$. With Jensen's inequality and a constant shift, (\ref{con_entropy}) can be approximated by \cite{Zeng_CC_approx}
\begin{equation}\label{con_entropy_approx}
\begin{split}       &H(\mathbf{s}|y,\mathbf{h}^H\mathbf{P})^\text{approx.}\\
    &=\frac{1}{|\mathcal{X}|} \sum_{m=1}^{|\mathcal{X}|} \log_2 \sum_{l=1}^{|\mathcal{X}|} \exp \Big( -\frac{|\mathbf{h}^H\mathbf{P} \mathbf{d}_{m,l}(\mathcal{X})|^2}{2\sigma^2} \Big).
\end{split}
\end{equation}

\subsection{Achievable Rate of RSMA}
\begin{figure*}[!t]
\hrule
    \begin{flalign}\label{CC_rate_common_full}
         R_{\text{c},k} =&\log_2|\mathcal{X}_\text{c}| - \frac{1}{|\mathcal{X}_\text{c}\times\mathcal{X}_\mathcal{K}|} \sum_{m=1}^{|\mathcal{X}_\text{c}\times\mathcal{X}_\mathcal{K}|} \mathbb{E}_n \Big\{ \log_2 \sum_{l=1}^{|\mathcal{X}_\text{c}\times\mathcal{X}_\mathcal{K}|} \exp \Big( -\frac{|\mathbf{d}_{m,l}(\mathbf{h}^H\mathbf{p}_\text{c}\mathcal{X}_\text{c}+\sum_{a\in\mathcal{K}}\mathbf{h}^H\mathbf{p}_a\mathcal{X}_a) + n|^2}{\sigma^2} \Big) \Big\}&&\nonumber\\
         &+\frac{1}{|\mathcal{X}_{\mathcal{K}}|} \sum_{m=1}^{|\mathcal{X}_{\mathcal{K}}|} \mathbb{E}_n \Big\{ \log_2 \sum_{l=1}^{|\mathcal{X}_{\mathcal{K}}|} \exp \Big( -\frac{|\mathbf{d}_{m,l}(\sum_{a\in\mathcal{K}}\mathbf{h}^H\mathbf{p}_a\mathcal{X}_a) + n|^2}{\sigma^2} \Big) \Big\}&&
    \end{flalign}

    \begin{flalign}\label{CC_rate_private_SIC_full}
         R_{\text{p},k}^{\text{SIC}} =&\log_2|\mathcal{X}_k| - \frac{1}{|\mathcal{X}_{\mathcal{K}}|} \sum_{m=1}^{|\mathcal{X}_{\mathcal{K}}|} \mathbb{E}_n \Big\{ \log_2 \sum_{l=1}^{|\mathcal{X}_{\mathcal{K}}|} \exp \Big( -\frac{|\mathbf{d}_{m,l}(\sum_{a\in\mathcal{K}}\mathbf{h}^H\mathbf{p}_a\mathcal{X}_a) + n|^2}{\sigma^2} \Big) \Big\}&&\nonumber\\
         &+\frac{1}{|\mathcal{X}_{\mathcal{K}/k}|} \sum_{m=1}^{|\mathcal{X}_{\mathcal{K}/k}|} \mathbb{E}_n \Big\{ \log_2 \sum_{l=1}^{|\mathcal{X}_{\mathcal{K}/k}|} \exp \Big( -\frac{|\mathbf{d}_{m,l}(\sum_{a\in\mathcal{K}/k}\mathbf{h}^H\mathbf{p}_a\mathcal{X}_a) + n|^2}{\sigma^2} \Big) \Big\}&&
    \end{flalign}

    \begin{flalign}\label{CC_rate_private_non_SIC_full}
         R_{\text{p},k}^{\text{SIC-free}} =&\log_2|\mathcal{X}_k| - \frac{1}{|\mathcal{X}_\text{c}\times\mathcal{X}_\mathcal{K}|} \sum_{m=1}^{|\mathcal{X}_\text{c}\times\mathcal{X}_\mathcal{K}|} \mathbb{E}_n \Big\{ \log_2 \sum_{l=1}^{|\mathcal{X}_\text{c}\times\mathcal{X}_\mathcal{K}|} \exp \Big( -\frac{|\mathbf{d}_{m,l}(\mathbf{h}^H\mathbf{p}_\text{c}\mathcal{X}_\text{c}+\sum_{a\in\mathcal{K}}\mathbf{h}^H\mathbf{p}_a\mathcal{X}_a) + n|^2}{\sigma^2} \Big) \Big\}&&\nonumber\\
         &+\frac{1}{|\mathcal{X}_\text{c}\times\mathcal{X}_{\mathcal{K}/k}|} \sum_{m=1}^{|\mathcal{X}_\text{c}\times\mathcal{X}_{\mathcal{K}/k}|} \mathbb{E}_n \Big\{ \log_2 \sum_{l=1}^{|\mathcal{X}_\text{c}\times\mathcal{X}_{\mathcal{K}/k}|} \exp \Big( -\frac{|\mathbf{d}_{m,l}(\mathbf{h}^H\mathbf{p}_\text{c}\mathcal{X}_\text{c} + \sum_{a\in\mathcal{K}/k}\mathbf{h}^H\mathbf{p}_a\mathcal{X}_a) + n|^2}{\sigma^2} \Big) \Big\}&&
    \end{flalign}
\hrule
\end{figure*}

Returning to the RSMA system model described in Section \ref{Section_system_model}, the achievable rate of the common stream at user-$k$, $k\in\mathcal{K}$, can be derived as \cite{Sibo2024}
\begin{equation}\label{CC_rate_common_1}
\begin{split}
    R_{\text{c},k}
    =& I(s_\text{c};y_k|\mathbf{h}_k^H\mathbf{P})\\
    =& I\Big(\{s_\text{c},s_1,...,s_K\} ; y_k \Big| \mathbf{h}_k^H\mathbf{P}\Big)\\
    &- I\Big(\{s_1,...,s_K\} ; y_k \Big| \mathbf{h}_k^H\mathbf{P},s_\text{c}\Big)\\
    =& H(s_\text{c},s_1,...,s_K) - H\Big(s_\text{c},s_1,...,s_K \Big| y_k,\mathbf{h}_k^H\mathbf{P}\Big)\\
    &- H(s_1,...,s_K) + H\Big(s_1,...,s_K \Big| y_k,\mathbf{h}_k^H\mathbf{P},s_\text{c}\Big),
\end{split}
\end{equation}
where the expansion is made from the chain rule of mutual information \cite{Cover}. To ensure that the common stream can be decoded by all the users, the common stream rate depends on the worst-case user, i.e.,
\begin{equation}\label{common_rate}
    R_\text{c} = \min_k R_{\text{c},k}.
\end{equation}
Similarly, for the private stream intended for user-$k$, if the receiver adopts SIC implementation, the achievable rate is expressed as
\begin{equation}\label{CC_rate_private_SIC}
\begin{split}
    &R_{\text{p},k}^{\text{SIC}}\\
    &= I \Big(s_k ; y_k \Big| \mathbf{h}_k^H\mathbf{P},s_\text{c} \Big)\\
    &= H(\{s_k|k\in\mathcal{K}\}) - H \Big(\{s_k|k\in\mathcal{K}\} \Big| y_k - \mathbf{h}_k^H\mathbf{p}_\text{c} s_\text{c},\mathbf{h}_k^H\mathbf{P} \Big)\\
    &\;\;\;\; -H(\{s_{k'}|k'\in\mathcal{K}/k\})\\
    &\;\;\;\; + H \Big(\{s_{k'}|k'\in\mathcal{K}/k\} \Big| y_k - \mathbf{h}_k^H\mathbf{p}_\text{c} s_\text{c}-\mathbf{h}_k^H\mathbf{p}_ks_k,\mathbf{h}_k^H\mathbf{P} \Big).
\end{split}
\end{equation}
With SIC-free receivers, the achievable rate is expressed as
\begin{equation}\label{CC_rate_private_no_SIC}
\begin{split}
    &R_{\text{p},k}^{\text{SIC-free}}\\
    &= I \Big(s_k ; y_k \Big| \mathbf{h}_k^H\mathbf{P} \Big)\\
    &= H(s_\text{c}, \{s_k|k\in\mathcal{K}\}) - H \Big(s_\text{c}, \{s_k|k\in\mathcal{K}\} \Big| y_k,\mathbf{h}_k^H\mathbf{P} \Big)\\
     &\;\;\;\;- H(s_\text{c},\{s_{k'}|k'\in\mathcal{K}/k\})\\
     &\;\;\;\; + H \Big(s_\text{c},\{s_{k'}|k'\in\mathcal{K}/k\} \Big| y_k-\mathbf{h}_k^H\mathbf{p}_ks_k,\mathbf{h}_k^H\mathbf{P} \Big).
\end{split}
\end{equation}
Note that the difference between (\ref{CC_rate_private_SIC}) and (\ref{CC_rate_private_no_SIC}) lies in the availability of knowledge on $s_\text{c}$.
The full expressions of $R_{\text{c},k}$,  $R_{\text{p},k}^{\text{SIC}}$ and $R_{\text{p},k}^{\text{SIC-free}}$ are given by (\ref{CC_rate_common_full})-(\ref{CC_rate_private_non_SIC_full}) respectively, where $\mathcal{X}_\text{c}$ is the constellation set used for the common stream, $\mathcal{X}_{\text{p},k}$ is the constellation set used for the private stream intended for the $k$-th user, and $\mathcal{X}_\mathcal{M} \triangleq \prod_{k\in\mathcal{M}}\mathcal{X}_{\text{p},k}$ for certain set $\mathcal{M}$. The approximations of $R_{\text{c},k}$, $R_{\text{p},k}^{\text{SIC}}$ and $R_{\text{p},k}^{\text{SIC-free}}$ are given by (\ref{CC_rate_common_approx})-(\ref{CC_rate_private_non_SIC_approx}) respectively.

\begin{figure*}[!t]
    \hrule
    \begin{flalign}\label{CC_rate_common_approx}
         R_{\text{c},k}^\text{approx.} =&\log_2|\mathcal{X}_\text{c}| - \frac{1}{|\mathcal{X}_\text{c}\times\mathcal{X}_\mathcal{K}|} \sum_{m=1}^{|\mathcal{X}_\text{c}\times\mathcal{X}_\mathcal{K}|} \log_2 \sum_{l=1}^{|\mathcal{X}_\text{c}\times\mathcal{X}_\mathcal{K}|} \exp \Big( -\frac{|\mathbf{d}_{m,l}(\mathbf{h}^H\mathbf{p}_\text{c}\mathcal{X}_\text{c}+\sum_{a\in\mathcal{K}}\mathbf{h}^H\mathbf{p}_a\mathcal{X}_a)|^2}{2\sigma^2} \Big)&&\nonumber\\
         &+\frac{1}{|\mathcal{X}_{\mathcal{K}}|} \sum_{m=1}^{|\mathcal{X}_{\mathcal{K}}|} \log_2 \sum_{l=1}^{|\mathcal{X}_{\mathcal{K}}|} \exp \Big( -\frac{|\mathbf{d}_{m,l}(\sum_{a\in\mathcal{K}}\mathbf{h}^H\mathbf{p}_a\mathcal{X}_a)|^2}{2\sigma^2} \Big) &&
    \end{flalign}

    \begin{flalign}\label{CC_rate_private_SIC_approx}
         R_{\text{p},k}^{\text{SIC, approx.}} =&\log_2|\mathcal{X}_k| - \frac{1}{|\mathcal{X}_{\mathcal{K}}|} \sum_{m=1}^{|\mathcal{X}_{\mathcal{K}}|} \log_2 \sum_{l=1}^{|\mathcal{X}_{\mathcal{K}}|} \exp \Big( -\frac{|\mathbf{d}_{m,l}(\sum_{a\in\mathcal{K}}\mathbf{h}^H\mathbf{p}_a\mathcal{X}_a)|^2}{2\sigma^2} \Big)&&\nonumber\\
         &+\frac{1}{|\mathcal{X}_{\mathcal{K}/k}|} \sum_{m=1}^{|\mathcal{X}_{\mathcal{K}/k}|} \log_2 \sum_{l=1}^{|\mathcal{X}_{\mathcal{K}/k}|} \exp \Big( -\frac{|\mathbf{d}_{m,l}(\sum_{a\in\mathcal{K}/k}\mathbf{h}^H\mathbf{p}_a\mathcal{X}_a)|^2}{2\sigma^2} \Big) &&
    \end{flalign}

    \begin{flalign}\label{CC_rate_private_non_SIC_approx}
         R_{\text{p},k}^{\text{SIC-free, approx.}} =&\log_2|\mathcal{X}_k| - \frac{1}{|\mathcal{X}_\text{c}\times\mathcal{X}_\mathcal{K}|} \sum_{m=1}^{|\mathcal{X}_\text{c}\times\mathcal{X}_\mathcal{K}|} \log_2 \sum_{l=1}^{|\mathcal{X}_\text{c}\times\mathcal{X}_\mathcal{K}|} \exp \Big( -\frac{|\mathbf{d}_{m,l}(\mathbf{h}^H\mathbf{p}_\text{c}\mathcal{X}_\text{c}+\sum_{a\in\mathcal{K}}\mathbf{h}^H\mathbf{p}_a\mathcal{X}_a)|^2}{2\sigma^2} \Big)&&\nonumber\\
         &+\frac{1}{|\mathcal{X}_\text{c}\times\mathcal{X}_{\mathcal{K}/k}|} \sum_{m=1}^{|\mathcal{X}_\text{c}\times\mathcal{X}_{\mathcal{K}/k}|} \log_2 \sum_{l=1}^{|\mathcal{X}_\text{c}\times\mathcal{X}_{\mathcal{K}/k}|} \exp \Big( -\frac{|\mathbf{d}_{m,l}(\mathbf{h}^H\mathbf{p}_\text{c}\mathcal{X}_\text{c}+\sum_{a\in\mathcal{K}/k}\mathbf{h}^H\mathbf{p}_a\mathcal{X}_a)|^2}{2\sigma^2} \Big)&&
    \end{flalign}
    \hrule
 \end{figure*}

Because $s_\text{c}$ carries common part messages for potentially multiple users, $R_\text{c}$ is distributed among these users, and this distribution depends on the
common part message each user contributed to $W_\text{c}$. Let $C_k$ denote the portion of the common stream rate that is allocated to user-$k$. Inherently, $C_k \geq 0$, $\forall k$, and $\sum_k C_k = R_\text{c}$. The rate experienced by user-$k$ is given by\footnote{Hereinafter, the superscript "SIC/SIC-free" indicates that the expression depends on whether SIC receivers or SIC-free receivers are used.}
\begin{equation}
    R_k^{\text{SIC/SIC-free}} = C_k + R_{\text{p},k}^{\text{SIC/SIC-free}}.
\end{equation}

The performance loss incurred by removing SIC receiver at user-$k$ can be characterized as
\begin{equation}
\begin{split}
    &R_k^{\text{SIC}} - R_k^{\text{SIC-free}}\\
    =& I(s_k;y_k) - I(s_k;y_k|s_\text{c})\\
    =& I(s_k;\sum_{k'\in\mathcal{K}} \mathbf{h}_k^H\mathbf{p}_{k'} s_{k'} + \mathbf{h}_k^H\mathbf{p}_\text{c} s_\text{c} + n_k)\\
    &- I(s_k;\sum_{k'\in\mathcal{K}}\mathbf{h}_k^H\mathbf{p}_{k'} s_{k'} +n_k).
\end{split}
\end{equation}
Understanding this difference requires addressing the following question: Consider a channel that takes $\mathbf{h}_k^H\mathbf{p}_ks_k$ as input and is disturbed by interference and noise denoted by $\sum_{k'\in\mathcal{K}/k} \mathbf{h}_k^H\mathbf{p}_{k'} s_{k'} + n_k$. How detrimental is it to the achievable rate if an additional interference, $\mathbf{h}_k^H\mathbf{p}_\text{c}s_\text{c}$, is added?
It is intuitive that $s_\text{c}$ is less detrimental to the achievable rate if it is from finite alphabet rather than Gaussian distributed, which motivated our analysis of SIC-free RSMA under finite alphabets. A recent work \cite{Sibo_FAGCI} addresses the impact of finite alphabet interference and shows that this impact is non-monotonic w.r.t. the interference power and can be low even if the interference is strong. Specifically, \cite{Sibo_FAGCI} shows that
\begin{equation}
\begin{split}
    &\lim_{\mathrm{E}[|\mathbf{h}_k^H\mathbf{p}_\text{c}|^2]\rightarrow+\infty} I(s_k;\sum_{k'\in\mathcal{K}} \mathbf{h}_k^H\mathbf{p}_{k'} s_{k'} + \mathbf{h}_k^H\mathbf{p}_\text{c} s_\text{c} + n_k)\\
    = &I(s_k;\sum_{k'\in\mathcal{K}} \mathbf{h}_k^H\mathbf{p}_{k'} s_{k'} + n_k).
\end{split}
\end{equation}
This indicates that the achievable rate of SIC-free RSMA can be made similar to SIC-based RSMA through proper precoding design.
\section{Precoding Optimization}\label{Section_precoding_opt}
In this section, we consider WSR and MMF maximization problems for RSMA with SIC and SIC-free receivers, and propose precoding optimization algorithms to address them.
\subsection{Weighted Sum-Rate Optimization}
The WSR problem can be expressed as
\begin{equation}\label{WSR}
\begin{split}
    \mathcal{P}_1: \;\; \underset{\mathbf{P},\mathbf{c}}{\max} \: \: & \sum_{k\in\mathcal{K}} u_k R_{k}^{\text{SIC/SIC-free}}\\ 
    \text{s.t.} \: \: \: \;
    & \|\mathbf{P}\|_\text{F}^2 \leq P_\text{T}\\
    & \sum_{k\in\mathcal{K}} C_k = R_\text{c}\\
    & \mathbf{c} \succcurlyeq \mathbf{0}^{1\times K},
\end{split}
\end{equation}
where $u_k$, $k\in\mathcal{K}$, denotes the weight assigned to user-$k$, $P_\text{T}$ denotes the transmit power budget, and $\mathbf{c} = [C_1,C_2,...,C_K]$.
\begin{proposition}\label{WSR_prop}
    Let $i \in \arg \max_{k \in \mathcal{K}} u_k$. If $(\mathbf{P}^\star, \mathbf{c}^\star)$ is a solution of $\mathcal{P}_1$, $(\mathbf{P}^\star, \mathbf{c}^{\star'})$ is another solution where $\mathbf{c}^{\star'} = [C_1^{\star'},C_2^{\star'},...,C_K^{\star'}]$ satisfies $C_i^{\star'} = R_\text{c}$ and $C_j^{\star'} = 0, \forall j \in \mathcal{K} \setminus i$.
\end{proposition}
\begin{IEEEproof}
    See Appendix \ref{proof of WSR_prop}.
\end{IEEEproof}

\begin{remark}
    The intuition behind Proposition \ref{WSR_prop} is of significant value for the practical implementation of RSMA, as it implies that when the objective is to maximize the WSR, only splitting the message of the most weighted users (or one of the most weighted users if multiple users are most weighted) is no worse than splitting multiple user messages. In practice, the splitting of messages requires additional implementation and control signaling, making splitting only one message a more favorable option.
\end{remark}

\begin{remark}
    The proof of Proposition \ref{WSR_prop} does not assume any input distribution, which means that Proposition \ref{WSR_prop} holds for arbitrary input distribution, including Gaussian input, and any other distribution for discrete or continuous space.
\end{remark}

As a direct consequence of Proposition \ref{WSR_prop}, one solution of $\mathbf{c}$ can already be determined given $u_k$, $\forall k\in\mathcal{K}$. As a result, $\mathbf{c}$ can be removed from the design variables, and $\mathcal{P}_1$ is simplified into
\begin{equation}\label{WSR_2}
\begin{split}
    \mathcal{P}_2: \;\; \underset{\mathbf{P}}{\max} \: \: & u_{k'} R_\text{c} + \sum_{k\in\mathcal{K}} u_{k'} R_{\text{p},k}^{\text{SIC/SIC-free}}\\ 
    \text{s.t.} \: \: \: \;
    & \|\mathbf{P}\|_\text{F}^2 \leq P_\text{T},
\end{split}
\end{equation}
where $k' \in \arg \max_{k \in \mathcal{K}} u_k$.

One challenge in solving $\mathcal{P}_2$ lies in the fact that the expectation in (\ref{con_entropy}) cannot be reduced to an analytical form. One way to address it is to estimate (\ref{con_entropy}) empirically and optimize that estimate, but this can introduce a significant computational burden to algorithms. We adapt an alternative approach by using the approximation given by (\ref{CC_rate_common_approx})-(\ref{CC_rate_private_non_SIC_approx}) during the optimization process. The effectiveness of this approach has been illustrated, for example, in \cite{Zeng_CC_approx, Pei_MAC_channel, Sibo2024,Jin_GQMP}. To tackle the non-convexity and complicated structure of (\ref{con_entropy_approx}) and the discontinuity brought about by the point-wise minimum in (\ref{common_rate}), subgradient method \cite{Rockafellar} is chosen to solve $\mathcal{P}_2$. 

The gradient of (\ref{con_entropy_approx}) is given by (\ref{con_entropy_approx_gradient}),
\begin{figure*}
\begin{equation}\label{con_entropy_approx_gradient}
    \nabla_\mathbf{P} H(\mathbf{s}|y,\mathbf{h}^H\mathbf{P})^\text{approx.}
    = \frac{\log_2 e}{|\mathcal{X}|}\sum_{m=1}^{|\mathcal{X}|} \frac{1}{\sum_{l=1}^{|\mathcal{X}|} u(\mathbf{P})}\sum_{l=1}^{|\mathcal{X}|} u(\mathbf{P}) \left( -\frac{\mathbf{hh}^H\mathbf{P}\mathbf{d}_{m,l}(\mathcal{X})\mathbf{d}_{m,l}(\mathcal{X})^H}{2\sigma^2} \right)
\end{equation}
\end{figure*}
where 
\begin{equation}
    u(\mathbf{P}) = \sum_{l=1}^{|\mathcal{X}|} \exp \Big( -\frac{|\mathbf{h}^H\mathbf{P} \mathbf{d}_{m,l}(\mathcal{X})|^2}{2\sigma^2} \Big).
\end{equation}
Based on (\ref{con_entropy_approx_gradient}), $\nabla_\mathbf{P}R_{\text{c},k}$, $\nabla_\mathbf{P}R_{k}^{\text{SIC}}$ and $\nabla_\mathbf{P}R_{k}^{\text{SIC-free}}$ can be obtained trivially. One subgradient\footnote{In this work, we use Clarke subdifferential for non-convex non-smooth functions \cite{clarke1990optimization}.} of the WSR can be obtained as
\begin{equation}\label{subgrad_WSR}
    g(R_{\text{WSR}}^{\text{SIC/SIC-free}}) =  u_{k'}\nabla_\mathbf{P}R_{\text{c},k^{\circ}} + \sum_{k\in\mathcal{K}} u_k\nabla_\mathbf{P}R_{k}^{\text{SIC/SIC-free}},
\end{equation}
where $k^{\circ} = \arg \min_{k\in\mathcal{K}} R_{\text{c},k}$.

It is easy to see that any solution of $\mathcal{P}_2$, denoted by $\mathbf{P}^\star$, must satisfy $\|\mathbf{P}^\star\|_\text{F}^2 = P_\text{T}$. We therefore use the projection of $\mathbf{P}$ in $\{\mathbf{P} \big| \|\mathbf{P}\|_\text{F}^2 = P_\text{T}\}$ as the updated solution after every iteration. The proposed Projected SubGradient Ascent (PSGA) algorithm is summarized in Algorithm \ref{SR_opt_alg}\footnote{In Algorithm \ref{SR_opt_alg}, depending on the assumption about receivers, $R_\text{WSR} = R_{\text{WSR}}^{\text{SIC}} \text{ or } R_{\text{WSR}}^{\text{SIC-free}}$.}. For choosing the step length, $t$, we use a modified backtracking line search specific for PSGA as described in Algorithm \ref{Modified_backtracking}. Compared to the classical backtracking for gradient descent/ascent, the proposed backtracking method incorporates projection when evaluating the predicted objective value after updating, ensuring a non-decreasing sequence of objectives. The convergence of general projected gradient methods is discussed in \cite{Luenberger}, although the convergence rate of such methods for non-convex problems is generally difficult to analyze. Note that each iteration in Algorithm \ref{Modified_backtracking} leads to a non-decreasing variation of the objective function, and the objective function of (\ref{WSR}) is upper-bounded. Therefore, Algorithm \ref{SR_opt_alg} is guaranteed to converge. The convergence is also numerically evaluated in Figure \ref{SR_Convergence}, where a monotonic increase in the objective function and the convergence can be observed.

\begin{algorithm}[!t]
    \caption{Projected Subgradient Ascent for Weight Sum-rate Maximization}
    \label{SR_opt_alg}
    \begin{algorithmic}[1]
        \REQUIRE $P_\text{T}$, $\mathcal{X}_\text{c}$, $\mathcal{X}_k$, $\mathbf{h}_k$, $k\in\mathcal{K}$, $\epsilon$, $v_\text{max}$.
        \ENSURE $\mathbf{P}^\star$
        \STATE Initialize $\mathbf{P}^0$.
        \STATE $v:=0$.
        \STATE Update. $R_{\text{WSR}}^{v} := R_{\text{WSR}}(\mathbf{P}^{v})$.
        \REPEAT
            \STATE $\Delta\mathbf{P} := g( R_{\text{WSR}})$ according to (\ref{subgrad_WSR}).
            \STATE Choose step size $t$ via Algorithm \ref{Modified_backtracking}.
            \STATE Update. $\mathbf{P}^{v+1} := \mathbf{P}^v + t \Delta\mathbf{P}$.
            \STATE Projection. $\mathbf{P}^{v+1} := \frac{\sqrt{P_\text{T}}}{\|\mathbf{P}^{v+1}\|_\text{F}}\mathbf{P}^{v+1}$.
            \STATE Update. $R_\text{WSR}^{v+1} := R_\text{WSR}(\mathbf{P}^{v+1})$.
            \STATE $v:=v+1$.
        \UNTIL $\mid R_\text{WSR}^{v} - R_\text{WSR}^{v-1} \mid\; < \epsilon$ or $v \geq v_\text{max}$.
        \RETURN $\mathbf{P}^\star := \mathbf{P}^{v}$.
    \end{algorithmic}
\end{algorithm}

\begin{algorithm}[!t]
    \caption{Backtracking Line Search for PSGA}
    \label{Modified_backtracking}
    \begin{algorithmic}[1]
        \REQUIRE $\Delta\mathbf{P}$, $f(\cdot)$\footnotemark, $P_\text{T}$, $\alpha\in(0,0.5)$, $\beta\in(0,1)$ and $t_\text{min}$.
        \ENSURE $t$
        \STATE $t:=1$.
        \WHILE{$f(\frac{\sqrt{P_\text{T}}}{\|\mathbf{P}+t\Delta\mathbf{P}\|_\text{F}}(\mathbf{P}+t\Delta\mathbf{P})) \leq f(\mathbf{P}) + \alpha t\|\Delta\mathbf{P}\|^2_\text{F}$ \AND $t>t_\text{min}$,}
        \STATE{$t := \beta t$.}
        \ENDWHILE
        \IF{$t \leq t_\text{min}$,}
        \STATE {$t:=0$.}
        \ENDIF
        \RETURN $t$
    \end{algorithmic}   
\end{algorithm}

\footnotetext{$f(\cdot)$ refers to the objective function of interest. It is kept general here, as Algorithm \ref{Modified_backtracking} will be reused later to solve the MMF problem.}

\subsection{Max-Min Fairness Optimization}
We now consider an MMF problem as follows:
\begin{equation}\label{Max_min}
\begin{split}
    \mathcal{P}_3: \;\; \underset{\mathbf{P},\mathbf{c}}{\max} \: \: & \min_{k\in\mathcal{K}}\: \: R_k^\text{SIC/SIC-free}\\ 
    \text{s.t.} \: \: \: \;
    & \|\mathbf{P}\|_\text{F}^2 \leq P_\text{T}\\
    & \sum_{k\in\mathcal{K}} C_k = R_\text{c}\\
    & \mathbf{c} \succcurlyeq \mathbf{0}^{1\times K}.
\end{split}
\end{equation}

Note that $\mathcal{P}_3$ is also a non-smooth non-convex problem and hence is difficult to solve. To develop efficient algorithms for $\mathcal{P}_3$, we decouple it into the following two sub-problems,
\begin{equation}\label{Max_min_sub_c}
\begin{split}
    \mathcal{P}_4: \;\; \underset{\mathbf{c}}{\text{max}} \: \: & \min_{k\in\mathcal{K}}\: \: R_k(\mathbf{c};\mathbf{P})\\ 
    \text{s.t.} \: \: \: \;
    & \sum_{k\in\mathcal{K}} C_k = R_\text{c}\\
    & \mathbf{c} \succcurlyeq \mathbf{0}^{1\times K},
\end{split}
\end{equation}
and
\begin{equation}\label{Max_min_sub_P}
\begin{split}
    \mathcal{P}_5: \;\; \underset{\mathbf{P}}{\text{max}} \: \: & \min_{k\in\mathcal{K}}\: \: R_k(\mathbf{P};\mathbf{c})\\ 
    \text{s.t.} \: \: \: \;
    & \|\mathbf{P}\|_\text{F}^2 \leq P_\text{T},
\end{split}
\end{equation}
where $R_k(\mathbf{c};\mathbf{P})$ indicates the case where 
$R_k^\text{SIC/SIC-free}$ is determined with 
$\mathbf{P}$ as a given parameter and 
$\mathbf{c}$ as a variable, whereas $R_k(\mathbf{P};\mathbf{c})$ indicates the case in which $\mathbf{c}$ is given and $\mathbf{P}$ is the variable. Hence, $\mathcal{P}_4$ optimizes $\mathbf{c}$ with a fixed $\mathbf{P}$ while $\mathcal{P}_5$ optimizes $\mathbf{P}$ with a given $\mathbf{c}$. 

Note that $\mathcal{P}_4$ is linear programming; therefore, its global optimal solution can be efficiently obtained. In particular, we developed a closed-form solution as in Proposition \ref{c_opt_prop}.

\begin{proposition}\label{c_opt_prop}
    The global optimal solution of $\mathcal{P}_4$ is given in closed form by Algorithm \ref{c_opt alg}.
\end{proposition}
\begin{IEEEproof}
    See Appendix \ref{proof of c_opt_prop}.
\end{IEEEproof}

\begin{algorithm}[!t]
    \caption{Global Optimal Common Stream Allocation}
    \label{c_opt alg}
    \begin{algorithmic}[1]
        \REQUIRE $R_\text{c}$, $R_{\text{p},k}$, $k\in\mathcal{K}$.
        \ENSURE $\mathbf{c}^\star$
        \STATE $k':=K+1$.
        \REPEAT 
            \STATE $k':=k'-1$.
            \STATE $\mathbf{d} :=[R_{\text{p},1},...,R_{\text{p},k'}]$.
            \STATE $\eta := \frac{R_\text{c}+\mathbf{d}\mathbf{1}^{k'\times1}}{k'}$.
            \STATE $\mathbf{c}' := \eta \mathbf{1}^{k'\times1} -\mathbf{d}$.
        \UNTIL $\mathbf{c}'\succcurlyeq 0$.
        \RETURN $\mathbf{c}^\star := [\mathbf{c}',\; \mathbf{0}^{1\times(K-k')}]$.
    \end{algorithmic}
\end{algorithm}

To remove the discontinuity caused by the point-wise minimum in $\mathcal{P}_5$, we apply a log-sum-exp approximation to its objective function and consider $\mathcal{P}_6$.
\begin{equation}\label{Max_min_LSE}
\begin{split}
    \mathcal{P}_6: \;\; \underset{\mathbf{P}}{\text{max}} \: \: & \frac{1}{\gamma}\log \sum_{k\in\mathcal{K}} \exp \big(\gamma R_k(\mathbf{P};\mathbf{c})\big)\\ 
    \text{s.t.} \: \: \: \;
    & \|\mathbf{P}\|_\text{F}^2 \leq P_\text{T}.
\end{split}
\end{equation}
 The gap between point-wise minimum and its log-sum-exp approximation is characterized by the following \cite{Chen_2013}.
 \begin{equation}
     0\leq \min_{1\leq i \leq L} f_i - \frac{1}{\gamma}\log\sum_{i=1}^L\exp(\gamma f_i) \leq \frac{1}{\gamma}\log L,\;\;\; \gamma < 0
 \end{equation}
 
Let $R_\text{max-min}$ denote the objective function in (\ref{Max_min_LSE}). One subgradient of $R_\text{max-min}$ w.r.t. $\mathbf{P}$ is derived to be
\begin{equation}\label{subgrad_MM_1}
    g(R_\text{max-min}) = \frac{\sum_{k\in\mathcal{K}} \exp \big(\gamma R_k(\mathbf{P};\mathbf{c})\big) g\big(R_k(\mathbf{P};\mathbf{c})\big)}{\sum_{k\in\mathcal{K}} \exp \big(\gamma R_k(\mathbf{P};\mathbf{c})\big)},
\end{equation}
where
\begin{equation}\label{subgrad_MM_2}
    g\left(R_k(\mathbf{P};\mathbf{c})\right) = c_k\nabla_\mathbf{P} R_{\text{c},k^\circ} + \nabla_\mathbf{P} R_{\text{p},k},
\end{equation}
and $k^\circ= \arg \min_k R_{\text{c},k}$.
Similarly to $\mathcal{P}_2$, $\mathcal{P}_6$ can also be solved using PSGA. The solution of (\ref{Max_min}) is obtained by updating the solution for $\mathcal{P}_4$ and $\mathcal{P}_6$ alternatively, and the procedures are summarized in Algorithm \ref{GD for Max_min opt}. Note that, in Algorithm \ref{GD for Max_min opt}, the updates on $\mathbf{P}$  do not decrease the objective function due to the step sizes selected by Algorithm \ref{Modified_backtracking}, and the updates on $\mathbf{c}$ also do not decrease the objective function, because Algorithm \ref{c_opt alg} leads to the global optimum. Hence, every iteration in Algorithm \ref{GD for Max_min opt} leads to a non-decreasing variation of the objective function. It is trivial to see that the objective function of $\mathcal{P}_3$ is upper-bounded. Therefore, Algorithm \ref{GD for Max_min opt} is guaranteed to converge.

\begin{algorithm}[!t]
    \caption{Subgradient Ascent with Projection for MMF Optimization}
    \label{GD for Max_min opt}
    \begin{algorithmic}[1]
        \REQUIRE $P_\text{T}$, $\mathcal{X}_\text{c}$, $\mathcal{X}_k$, $\mathbf{h}_k$, $k\in\mathcal{K}$, $\gamma$, $\epsilon$, $v_\text{max}$.
        \ENSURE $\mathbf{P}^\star$ and $\mathbf{c}^\star$.
        \STATE Initialize $\mathbf{P}^0$.
            \STATE $v:=0$.
            \STATE Update. $R_\text{max-min}^{0} := R_\text{max-min}(\mathbf{P}^0)$.
            \REPEAT
                \STATE Update $R_\text{c}$, $R_{\text{p},k}$ based on $\mathbf{P}^v$.
                \STATE Update $\mathbf{c}$ using Algorithm \ref{c_opt alg}.
                \STATE $\Delta\mathbf{P} := g( R_\text{max-min})$ according to (\ref{subgrad_MM_1}) and (\ref{subgrad_MM_2}).
                \STATE Choose step size $t$ via Algorithm \ref{Modified_backtracking}.
                \STATE Update. $\mathbf{P}^{v+1} := \mathbf{P}^v + t \Delta\mathbf{P}$.
                \STATE Projection. $\mathbf{P}^{v+1} := \frac{\sqrt{P_\text{T}}}{\|\mathbf{P}^{v+1}\|_\text{F}}\mathbf{P}^{v+1}$.
                \STATE Update. $R_\text{max-min}^{v+1} := R_\text{max-min}(\mathbf{P}^{v+1})$.
                \STATE $v:=v+1$.
            \UNTIL $\mid R_\text{max-min}^{v} - R_\text{max-min}^{v-1} \mid\; < \epsilon$ or $v \geq v_\text{max}$.
        \RETURN $\mathbf{P}^\star := \mathbf{P}^{v}$, $\mathbf{c}^\star = \mathbf{c}$.
    \end{algorithmic}
\end{algorithm}
\section{Optimization for Large-Scale Systems}\label{Section_large_system}
In this section, we propose methods to extend the precoding designs introduced in Section \ref{Section_precoding_opt} to systems with a large number of transmit antennas and a large number of co-scheduled users.

To solve linear precoding design problems for large-scale systems, the inherent challenge is the growing size of the design variable, i.e., the precoding matrix, which leads to growing computational complexity. For problems considering finite constellations, the even more fatal issue is the exponential growth in the cardinality of observed constellations at each user. It can be observed from (\ref{con_entropy}) and (\ref{con_entropy_approx}) that the computational complexity grows exponentially with respect to the number of received signal sources. An estimate of computational complexity of the sum-rate of a MU-MIMO system with different number of users and constellation cardinality is given in Table \ref{Table: Complexity from FLOPs} in terms of floating point operations (FLOPs). Practically, evaluating scenarios with more than 12 bits transmitted over all the streams becomes difficult. 
\begin{table}
    \centering
    \caption{An estimate of the number of FLOPs for computing Constellation-Constrained Rate}
    \renewcommand{\arraystretch}{1.2}
    \begin{tabular}{|c|c|c|c|c|}
        \hline
        $K$ & QPSK & 16QAM & 64QAM \\
        \hline
        2 & $1.18\times10^8$ & $2.77\times10^{10}$ & $7.05\times10^{12}$\\
        \hline
        6 & $2.46\times10^{13}$ & $3.56\times10^{20}$ & $5.95\times10^{27}$\\
        \hline
        10 & $2.45\times10^{18}$ & $2.55\times10^{30}$ & $2.79\times10^{42}$\\
        \hline
        50 & $1.79\times10^{67}$ & $2.72\times10^{127}$ & $4.36\times10^{187}$\\
        \hline
    \end{tabular}
    \label{Table: Complexity from FLOPs}
    \vspace{0.6cm}
        \centering
    \caption{An estimate of the number of FLOPs for computing Constellation-Constrained Rate with user grouping and interference nulling}
    \renewcommand{\arraystretch}{1.2}
    \begin{tabular}{|c|c|c|c|c|}
        \hline
        $K$ & QPSK & 16QAM & 64QAM \\
        \hline
        2 & $1.18\times10^8$ & $2.77\times10^{10}$ & $7.05\times10^{12}$\\
        \hline
        6 & $3.55\times10^{8}$ & $8.31\times10^{10}$ & $2.11\times10^{13}$\\
        \hline
        10 & $5.91\times10^{8}$ & $1.38\times10^{11}$ & $3.52\times10^{13}$\\
        \hline
        50 & $2.96\times10^{9}$ & $6.92\times10^{11}$ & $1.76\times10^{14}$\\
        \hline
    \end{tabular}
    \label{Table: Complexity from FLOPs with null space projection}
\end{table}
\begin{figure}
    \centering
    \vspace{0.5cm}\includegraphics[width=0.99\linewidth]{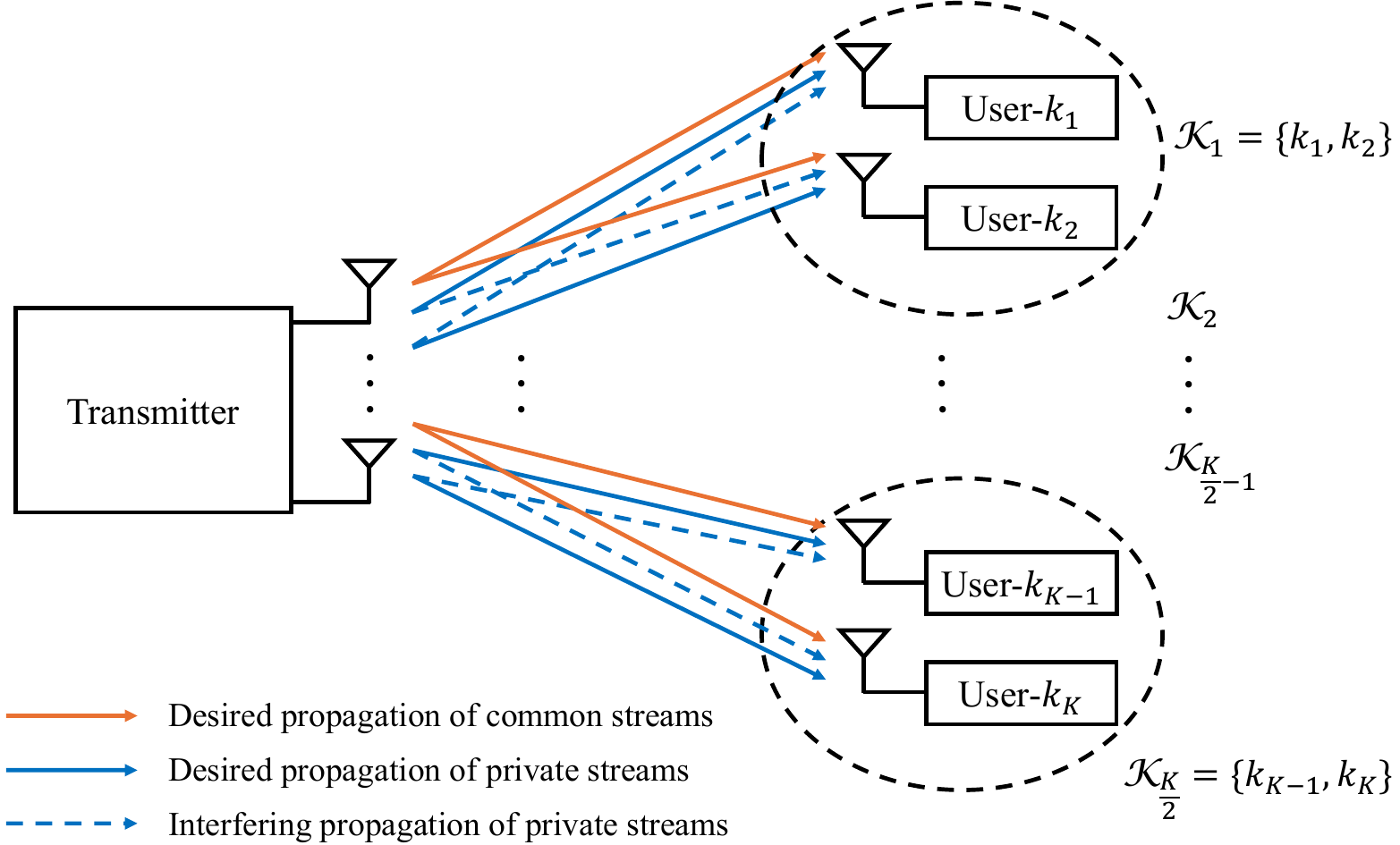}
    \caption{Illustration of user grouping.}
    \label{fig:user grouping}
\end{figure}

To enable precoding designs and evaluation for large-scale systems (e.g., massive MIMO) with practical computational burden and acceptable reduction in performance, we introduce user grouping and null space projection techniques. The idea is to partition the co-scheduled users into groups, enable one common stream for each group, and constrain the precoders such that inter-group interference is eliminated. Figure \ref{fig:user grouping} illustrates a system in which $K$ users (assuming that $K$ is even) are partitioned into $\frac{K}{2}$ groups, and each group contains two users. The elimination of inter-group interference is illustrated since signals carrying the common and private streams intended for a particular user do not propagate to other groups, although intra-group interference, i.e., interference caused by the private stream intended for a co-group user, potentially exists. In contrast to Table \ref{Table: Complexity from FLOPs}, an estimate of computational complexity with user grouping and interference nulling is given by Table \ref{Table: Complexity from FLOPs with null space projection}, where the growth in FLOPs is roughly linear w.r.t. the number of users.

Problem formulation involving user grouping and interference nulling can be written as the following:
\begin{subequations}
    \begin{align}
    \mathcal{P}_7: \;\; \underset{\mathbf{P},\mathbf{c},\mathcal{G}}{\max} \: \: & \Omega(\mathbf{P},\mathbf{c},\mathcal{G})\nonumber\\ 
    \text{s.t.} \;\;& \|\mathbf{P}\|_\text{F}^2 \leq P_\text{T}\\
    & |h_m^H\mathbf{p}_k|^2 = 0, \nonumber\\
    &\forall m \in \mathcal{K}_a, k\in \mathcal{K}_b, \mathcal{K}_a\in\mathcal{G}, \mathcal{K}_b\in\mathcal{G}, \mathcal{K}_a \neq \mathcal{K}_b \label{equ: int elimiation constraint 1}\\
    & |h_m^H\mathbf{p}_{\mathcal{K}_1}|^2 = 0, \forall m \notin {\mathcal{K}_a} \in \mathcal{G} \label{equ: int elimiation constraint 2}\\
    & \sum_{k\in\mathcal{K}_a} C_k = R_{\mathcal{K}_1}, \forall \mathcal{K}_a \in \mathcal{G} \\
    & \mathbf{c} \succcurlyeq \mathbf{0}^{1\times K},
    \end{align}
\end{subequations}
where 
\begin{equation}
    \Omega(\mathbf{P},\mathbf{c},\mathcal{G}) = \sum_{k\in\mathcal{K}} u_k R_k^{\text{SIC/SIC-free}}
\end{equation}
for the WSR problem, and 
\begin{equation}
    \Omega(\mathbf{P},\mathbf{c},\mathcal{G}) = \min_{k\in\mathcal{K}} R_k^{\text{SIC/SIC-free}}
\end{equation}
for the MMF problem. $\mathcal{G} = \{\mathcal{K}_1, \mathcal{K}_2, ...\}$ denotes the grouping, whose elements are sets of user indices allocated to different groups. $\mathbf{P} = [\mathbf{p}_{\mathcal{K}_1}, \mathbf{p}_{\mathcal{K}_2},...,\mathbf{p}_1, \mathbf{p}_2,..., \mathbf{p}_K]$, where $\mathbf{p}_{\mathcal{K}_i}$, $\forall \mathcal{K}_i \in \mathcal{G}$, is the common stream precoder intended for users in group $\mathcal{K}_i$. $R_{\mathcal{K}_i}$, $\forall \mathcal{K}_i \in \mathcal{G}$, denotes the achievable rate of the common stream intended for group $\mathcal{K}_i$. 
In addition to the constraints in $\mathcal{P}_1$ and $\mathcal{P}_3$, (\ref{equ: int elimiation constraint 1}) and (\ref{equ: int elimiation constraint 2}) are added to $\mathcal{P}_7$ to enforce inter-group interference nulling. As a result of (\ref{equ: int elimiation constraint 1}) and (\ref{equ: int elimiation constraint 2}), computation of the achievable rate of user-$k$, $\forall k\in \mathcal{K}$, can be simplified by only considering signals intended for users in the same group as user-$k$. Specifically, $\forall k\in\mathcal{K}_i$, $\forall \mathcal{K}_i \in \mathcal{G}$, the computation of (\ref{CC_rate_common_full}-\ref{CC_rate_private_non_SIC_full}) and (\ref{CC_rate_common_approx}-\ref{CC_rate_private_non_SIC_approx}) can be reduced by substituting $\mathcal{K}_i$ for $\mathcal{K}$ without affecting the results.

$\mathcal{P}_7$ involves designing $\mathcal{G}$, i.e., the user grouping, and therefore is a combinatorial optimization problem. The joint optimization of $\mathbf{P}$, $\mathbf{c}$ and $\mathcal{G}$ is difficult. We therefore design $\mathcal{G}$ through a heuristic approach and ensure (\ref{equ: int elimiation constraint 1}) and (\ref{equ: int elimiation constraint 2}) through multi-stage precoding, followed by optimizing $\mathbf{P}$ and $\mathbf{c}$. To address the growth in the dimension of $\mathbf{P}$ w.r.t. $n_\text{T}$, we also utilize a technique to reduce the dimensionality of the design variables to further accelerate optimization algorithms.

\subsection{User Grouping}
It is intuitive that, with an optimal precoding matrix, a user would suffer less interference from streams intended for co-scheduled users which have dissimilar channels. Hence, we heuristically group users with similar channels to limit the potential performance degradation. For the purpose of limiting the computational burden, the number of users within each group should be limited. We consider a special case where each group contains two users. The proposed user grouping algorithm is described in Algorithm \ref{Alg: User grouping}\footnote{We assume $K$ to be even for clarity, while extension to odd numbers is straightforward. For example, one can assume that one group contains only one user, while all other groups contain two users.}, where $q(\mathbf{h}_m,\mathbf{h}_n)$ is a predefined channel similarity metric. We use the cosine similarity of channels for our analysis, i.e.,
\begin{equation}
    q(\mathbf{h}_m,\mathbf{h}_n) = \frac{|\mathbf{h}_m^H\mathbf{h}_n|}{\|\mathbf{h}_m\|\|\mathbf{h}_n\|},
\end{equation}
although alternative metrics might provide superior results or serve different purposes.
\begin{algorithm}[!t]
    \caption{User Grouping Algorithm}
    \label{Alg: User grouping}
    \begin{algorithmic}[1]
        \REQUIRE $\mathbf{h}_k$, $k\in\mathcal{K}$.
        \ENSURE $\mathcal{G}$
        \STATE Initialize $\mathcal{G} = \varnothing$
        \STATE Compute $\mathbf{Q}\in\mathbb{C}^{K\times K}$, where\\ $[\mathbf{Q}]_{m,n}=q(\mathbf{h}_m,\mathbf{h}_n)$, $\forall m < n, m\in\mathcal{K}, n\in\mathcal{K}$, and\\
        $[\mathbf{Q}]_{m,n}=-1$, $\forall m \geq n, m\in\mathcal{K}, n\in\mathcal{K}$.
        \FOR{$l\leftarrow 1$ \TO $K/2$}
            \STATE $(a,b)\leftarrow\arg\max_{m,n} [\mathbf{Q}]_{m,n}$.
            \STATE Add $\{a,b\}$ to $\mathcal{G}$.
            \STATE $[\mathbf{Q}]_{m,n}:=-1$, $\forall (m,n) \in$ \\ $\{(m,n)|m=a\;\vee\; m=b\;\vee\; n=a\;\vee\; n=b\}$.
        \ENDFOR
        \RETURN $\mathcal{G}$
    \end{algorithmic}
\end{algorithm}
It is evident that Algorithm \ref{Alg: User grouping} can be extended to handle cases with more than two users per group.

\subsection{Null Space Projection and Dimensionality Reduction}
We propose three-stage precoding by decomposing each precoder into a product of two matrices and a vector, and design them sequentially. Suppose that the user grouping algorithm suggests $k\in\mathcal{K}_i$, where $\mathcal{K}_i$ denotes one of the groups. The common stream precoder for group $\mathcal{K}_i$ is denoted and decomposed as
\begin{equation}
    \mathbf{p}_{\mathcal{K}_i} = \mathbf{F}_{\mathcal{K}_i}\mathbf{G}_{\mathcal{K}_i}\mathbf{v}_{\mathcal{K}_i},
\end{equation}
and the private stream precoder for user-$k$ is decomposed as
\begin{equation}
    \mathbf{p}_{k} = \mathbf{F}_{\mathcal{K}_i}\mathbf{G}_{\mathcal{K}_i}\mathbf{v}_k,
\end{equation}
where $\mathbf{F}_{\mathcal{K}_i}$ projects the subsequent vector onto the null space of the channels of users outside $\mathcal{K}_i$, such that inter-group interference is eliminated. We denote and express the inter-group channels to users in $\mathcal{K}_i$ as
\begin{equation}
    \mathbf{H}_{\Bar{\mathcal{K}}_i} = [\mathbf{H}_{\mathcal{K}_1},...,\mathbf{H}_{\mathcal{K}_{i-1}},\mathbf{H}_{\mathcal{K}_{i+1}},...],
\end{equation}
where the columns of $\mathbf{H}_{\mathcal{K}_j}$ are the channel vectors of users in group $\mathcal{K}_j$. With eigenvalue decomposition (EVD), it is readily seen that inter-group interference nulling can be achieved by having 
\begin{equation}\label{1_stage_precoder}
    \mathbf{F}_{\mathcal{K}_i} = \Tilde{\mathbf{V}}_{\Bar{\mathcal{K}}_i},
\end{equation}
where $\Tilde{\mathbf{V}}_{\Bar{\mathcal{K}}_i}$ denotes the eigenvectors corresponding to the zero eigenvalues of $\mathbf{H}_{\Bar{\mathcal{K}}_i}\mathbf{H}_{\Bar{\mathcal{K}}_i}^H$. 

Apart from interference nulling, an additional benefit of $\mathbf{F}_{\mathcal{K}_i}$ is that the dimensionality of the subsequent design variable is reduced from $N_\text{T}$ to $N_\text{T}-K+2$, i.e., the dimensionality of the null space of inter-group channels. We introduce $\mathbf{G}_{\mathcal{K}_i}$ to further reduce the dimensionality of the design variable by extending the low-dimensional subspace property (LSP) introduced by \cite{Zhao_WMMSE} to finite constellation scenarios as follows. 
\begin{proposition}[Low-Dimensional Subspace Property]\label{LSP}
    Any non-trivial stationary point $\mathbf{P}^\star$ of $\mathcal{P}_1$ and $\mathcal{P}_3$ is in the range of $[\mathbf{h}_1,...,\mathbf{h}_K]$.
\end{proposition}

This is trivial to see as any components of $\mathbf{P}$ that lie outside the range of $[\mathbf{h}_1,...,\mathbf{h}_K]$ do not affect the objective function of $\mathcal{P}_1$ and $\mathcal{P}_3$, hence the proof of Proposition \ref{LSP} is omitted.

With inter-group interference nulling, the LSP implies that optimal $\mathbf{G}_{\mathcal{K}_i}\mathbf{v}_{\mathcal{K}_i}$ and $\mathbf{G}_{\mathcal{K}_i}\mathbf{v}_k$ lies in the column space of $\mathbf{F}_{\mathcal{K}_i}^H\mathbf{H}_{\mathcal{K}_j}$. This allows a reduction in the dimensionality in the design variable from $N_\text{T}-K+2$ to $2$ while ensuring that the optimal solution can still be achieved. Let 
\begin{equation}
\mathbf{F}_{\mathcal{K}_i}^H\mathbf{H}_{\mathcal{K}_i} = \mathbf{U}_{\mathcal{K}_i}\mathbf{\Sigma}\mathbf{V}_{\mathcal{K}_i}^H
\end{equation}
be compact singular value decomposition (SVD) of $\mathbf{F}_{\mathcal{K}_i}^H\mathbf{H}_{\mathcal{K}_j}$.
We restrict $\mathbf{G}_{\mathcal{K}_i}\mathbf{v}_k$ to be in the desired subspace by choosing\footnote{To achieve the same purpose, one can also have $\mathbf{G}_{\mathcal{K}_i} = \mathbf{F}_{\mathcal{K}_i}^H\mathbf{H}_{\mathcal{K}_j}$. However, it was observed numerically that the proposed $\mathbf{G}_{\mathcal{K}_i}$ leads to better convergence when optimizing $\mathbf{v}_k$. One potential reason is that the proposed $\mathbf{G}_{\mathcal{K}_i}$ guarantees that $\|\mathbf{p}_k\|^2 = \|\mathbf{v}_k\|^2$ and $\|\mathbf{p}_{\mathcal{K}_i,\text{c}}\|^2 = \|\mathbf{v}_{\mathcal{K}_i}\|^2$.}
\begin{equation}\label{2nd_stage_precoder}
    \mathbf{G}_{\mathcal{K}_i} = \mathbf{U}_{\mathcal{K}_i}\mathbf{I}\mathbf{V}_{\mathcal{K}_i}^H.
\end{equation}
The procedures of interference nulling and dimensionality reduction are summarized in Algorithm \ref{Alg: int nulling and dim reduction}. Given solutions for the first and second stages of precoding, the third stage, i.e., $\mathbf{v}_k$ and $\mathbf{v}_{\mathcal{K}_i}$, is designed in a manner similar to that in Section \ref{Section_precoding_opt}. Note that, as a result of inter-group interference nulling, inter-group interference does not need to be considered when computing the achievable rates and their gradients for $\mathcal{P}_1$ and $\mathcal{P}_3$. For the MMF problem, as each group posses a common stream, the design of common rate allocation, i.e., Algorithm \ref{c_opt alg}, is carried out for each group separately.

\begin{algorithm}[!t]
    \caption{Interference Nulling and Dimensionality Reduction}
    \label{Alg: int nulling and dim reduction}
    \begin{algorithmic}[1]
        \REQUIRE $\mathcal{G}$, $\mathbf{h}_k$, $k\in\mathcal{K}$.
        \ENSURE $\mathbf{F}_{\mathcal{K}_i}$ and $\mathbf{G}_{\mathcal{K}_i}$, $\forall \mathcal{K}_i \in \mathcal{G}$.
        \STATE Construct $\mathbf{H}_{\Bar{\mathcal{K}}_i}$, $\forall \mathcal{K}_i \in \mathcal{G}$.
        \STATE Perform EVD to $\mathbf{H}_{\Bar{\mathcal{K}}_i}\mathbf{H}_{\Bar{\mathcal{K}}_i}^H$ to obtain $\Tilde{\mathbf{V}}_{\Bar{\mathcal{K}}_i}$, $\forall \mathcal{K}_i \in \mathcal{G}$.
        \STATE $\mathbf{F}_{\mathcal{K}_i} := \Tilde{\mathbf{V}}_{\Bar{\mathcal{K}}_i}$, $\forall \mathcal{K}_i \in \mathcal{G}$.
        \STATE Perform SVD to $\mathbf{F}_{\mathcal{K}_i}^H\mathbf{H}_{\mathcal{K}_i}$ to obtain $\mathbf{U}_{\mathcal{K}_i}$ and $\mathbf{V}_{\mathcal{K}_i}$, $\forall \mathcal{K}_i \in \mathcal{G}$.
        \STATE $\mathbf{G}_{\mathcal{K}_i} := \mathbf{U}_{\mathcal{K}_i}\mathbf{I}\mathbf{V}_{\mathcal{K}_i}^H$.
        \RETURN $\mathbf{F}_{\mathcal{K}_i}$ and $\mathbf{G}_{\mathcal{K}_i}$, $\forall \mathcal{K}_i \in \mathcal{G}$.
    \end{algorithmic}
\end{algorithm}

\subsection{Complexity Analysis}
For clarity, we assume that all private streams apply the same alphabet, denoted by $\mathcal{X}$. For the general approaches proposed in Section \ref{Section_precoding_opt}, computing (\ref{CC_rate_common_approx}), (\ref{CC_rate_private_non_SIC_approx}) and their gradients w.r.t. $\mathbf{P}$ requires $\mathcal{O}(|\mathcal{X}_\text{c}|^2|\mathcal{X}|^{2K}Kn_\text{T})$ operations. Computing (\ref{CC_rate_private_SIC_approx}) and its gradient requires $\mathcal{O}(|\mathcal{X}|^{2K}Kn_\text{T})$. Therefore, the complexity of Algorithms \ref{SR_opt_alg} and \ref{GD for Max_min opt} is $\mathcal{O}(I_\text{it}|\mathcal{X}_\text{c}|^2|\mathcal{X}|^{2K}K^2n_\text{T})$, where $I_\text{it}$ is the number of iterations.

With user grouping, interference nulling and dimensionality reduction, $\mathcal{O}(|\mathcal{X}_\text{c}|^2|\mathcal{X}|^{2})$ is required for (\ref{CC_rate_common_approx}), (\ref{CC_rate_private_non_SIC_approx}) and their gradients. $\mathcal{O}(|\mathcal{X}|^{2})$ is required for (\ref{CC_rate_private_SIC_approx}) and its gradient. For user grouping, Algorithm \ref{Alg: User grouping} with the proposed similarity metric requires $\mathcal{O}(K^2N_\text{T})$. The complexity of EVD for an $n\times n$ matrix is $\mathcal{O}(n^3)$, and that of SVD for an $m\times n$ matrix is $\mathcal{O}(\min(mn^2, m^2n))$. Hence, Algorithm \ref{Alg: int nulling and dim reduction} requires $\mathcal{O}(n_\text{T}^3 + n_\text{T}^2K + \min(n_\text{T}K^2,n_\text{T}^2K))$. The overall complexity of user grouping, interference nulling, dimensionality reduction, and WSR/MMF optimization proposed in this section requires $\mathcal{O}(I_\text{it}|\mathcal{X}_\text{c}|^2|\mathcal{X}|^{2}K + n_\text{T}^3 + n_\text{T}^2K + n_\text{T}K^2)$. Compared to $\mathcal{O}(I_\text{it}|\mathcal{X}_\text{c}|^2|\mathcal{X}|^{2K}Kn_\text{T})$ for the general algorithm, the prohibitive exponent $K$ disappears, although there is an increase in the power of $n_\text{T}$ and $K$ due to EVD and SVD.

\section{Numerical Results}\label{Section_results}
This section introduces the simulation setup in use, including channel model and adaptive modulation, and presents numerical results on the performance of SIC-based and SIC-free RSMA.

\subsection{Channel Model}
Channels are modeled using Rician fading, i.e.,
\begin{equation}
    \mathbf{h}_k = \sqrt{\frac{\kappa_k}{\kappa_k+1}}\mathbf{h}_k^\text{LoS}(\theta_k^\text{az},\theta_k^\text{el}) + \sqrt{\frac{1}{\kappa_k+1}}\mathbf{h}_k^\text{NLoS},\forall k\in \mathcal{K},
\end{equation}
where $\mathbf{h}_k^\text{LoS}(\theta_k^\text{az},\theta_k^\text{el})$,  $\mathbf{h}_k^\text{NLoS}$ and $\kappa_k$ denote the Line-of-Sight (LoS) channel component, non-LoS channel component and and the Rician factor. $\mathbf{h}_k^\text{LoS}$ is characterized by the array geometry and the direction of departure in azimuth and elevation, denoted by $\theta_k^\text{az}$ and $\theta_k^\text{el}$ respectively. $\mathbf{h}_k^\text{NLoS}$ is modeled by Rayleigh fading with all the entries follow $\mathcal{CN}(0,1)$.

In this work, we aim to model channels in practical scenarios where users are in high-density demand areas, therefore simulating users in similar directions. For small-scale systems, a uniform linear array is assumed along the $y$ axis with $\theta_k^\text{el} = 0, \forall k\in \mathcal{K}$ since elevation does not affect the channels. For large-scale systems, a uniform rectangular array along the $y$-$z$ plane is assumed. It is assumed that $\kappa_1=...=\kappa_K=\kappa$.

\subsection{Adaptive Modulation}
The advantage of RSMA lies in its flexibility in adjusting the power and rates between common and private streams. To fully exploit this advantage under finite constellations, we allow RSMA to dynamically adjust the constellations. In particular, we enable RSMA to maximize the objective function by exhaustively searching for constellations from a pre-defined dictionary of transmission modes. To reduce the searching complexity, private streams are limited to using the same constellation in each transmission mode. To ensure fairness, different modes within the same dictionary are designed to produce the same maximum achievable rate, denoted by $R_\text{max}$. According to the above design principles, the four transmission mode dictionaries used in the simulations are documented in Table \ref{Table: Tx mode 2 UE 6 bits} - \ref{Table: Tx mode 3 UE 9 bits} in Appendix \ref{Appendix_Tx_mode}. Note that with Mode 1 in all the examples, RSMA is equivalent to SDMA, since the precoding optimization will allocate zero power to the common stream when the common stream constellation is empty. For large-scale systems, we assume that all the groups use the same transmission mode, which is searched from one of the pre-defined dictionaries. In the numerical results, where necessary, $K$ and $R_\text{max}$ are provided in the legends or captions to identify the transmission mode dictionary in use.

\subsection{Weighted Sum-Rate Performance}
\begin{figure*}[!h]
    \centering
    \subfigure[$N_T=2$, $K=2$, $\{\theta_1^\text{az}, \theta_2^\text{az}\}= \{0,\pi/18\}$, $\kappa=10\text{ dB}$.]{\includegraphics[width=7.2cm]{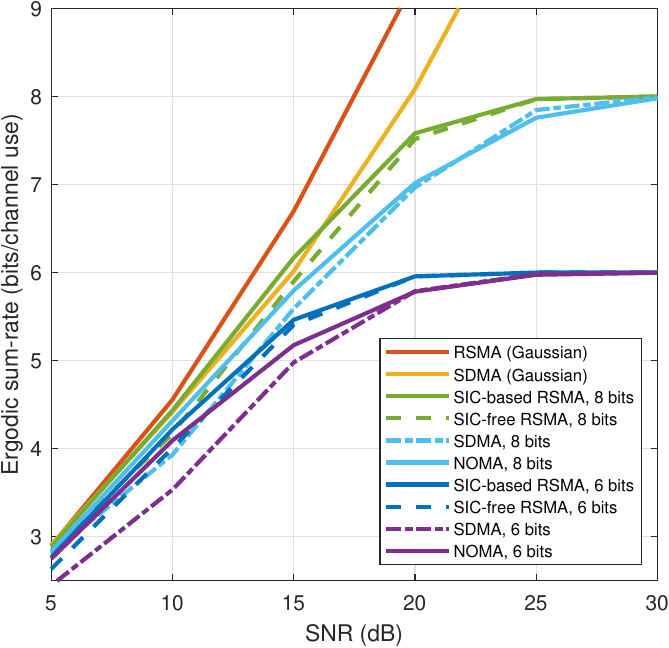}}\hspace{2cm}
    \subfigure[$N_T=4$, $K=3$, $\{\theta_1^\text{az}, \theta_2^\text{az}, \theta_3^\text{az}\}= \{-\pi/36,0,\pi/36\}$, $\kappa=10\text{ dB}$.]{\includegraphics[width=7.2cm]{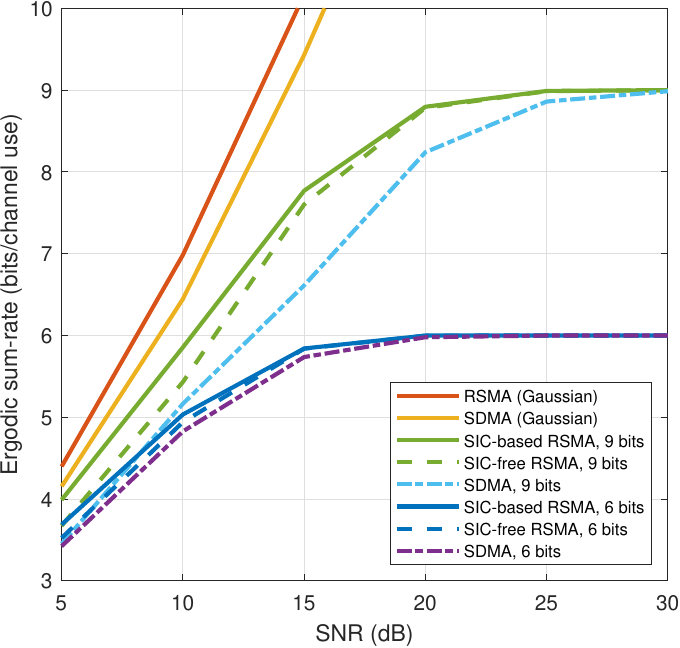}}
    \caption{Ergodic SR performance.}
    \label{SR}
\end{figure*}
We first evaluate the sum-rate performance by setting $[u_1,...,u_k] = \mathbf{1}^{1\times K}$. Figure \ref{SR} depicts the ergodic sum-rate of SIC-based and SIC-free RSMA under finite constellations. The performance of SDMA under finite constellations is also depicted by fixing the transmission mode to be the first one in each dictionaries. For two users, we also included power-domain Non-Orthogonal Multiple Access (NOMA) \cite{NOMA} as a benchmark, where both users apply the same constellation as in SDMA, and the user ordering is determined by the channel strength\footnote{For example, if $\|\mathbf{h}_1\|^2 \geq \|\mathbf{h}_2\|^2$, user-$1$ decodes user-$2$'s message, applies SIC and decodes its own message, whereas user-$2$ decodes its own message directly.}. Furthermore, the performance of RSMA and SDMA under Gaussian distributed input signals, following \cite{Mao_EURASIP}, is included as a reference. It can be observed that, in all cases, SIC-free RSMA preserves most of the superiority of SIC-based RSMA over SDMA and NOMA, which is consistent with the conclusion in \cite{Sibo2024}.

\begin{figure}
    \centering
    \includegraphics[width=7.2cm]{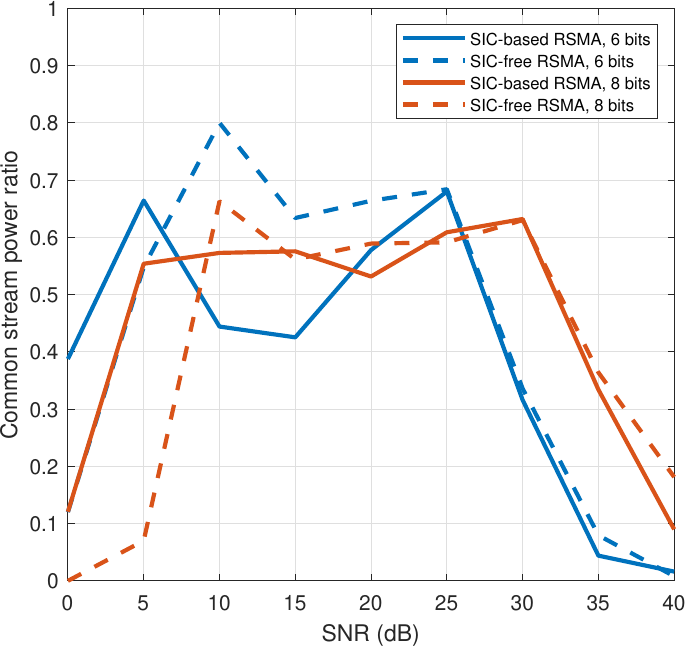}
    \caption{Ergodic common stream power ratio.}
    \label{SR_pow_allocation}
\end{figure}

Figure \ref{SR_pow_allocation} depicts the ratio of the average common stream power to the transmit power budget with SIC-based and SIC-free RSMA, i.e., $\mathbb{E}\left[\|\mathbf{p}_\text{c}\|^2/P_\text{T}\right]$, under the same setup as in Figure \ref{SR} (a). This result visualizes the fact that, despite SIC-based and SIC-free achieving similar performance, they demand very different precoder designs and power allocation strategies. Notably, SIC-free RSMA tends to allocate less power to the common stream in low SNR regime. This is to compensate for the fact that SIC-free receivers are not able to remove interference caused by the common stream. It can be observed that the algorithm tends to turn off the common stream at extremely high and low SNRs. It can also be observed that as $R_\text{max}$ increases, the power allocation curves shift to the right-hand side, because a higher achievable rate needs to be supported by a higher SNR.

\begin{figure}
    \centering
    \includegraphics[width=7.2cm]{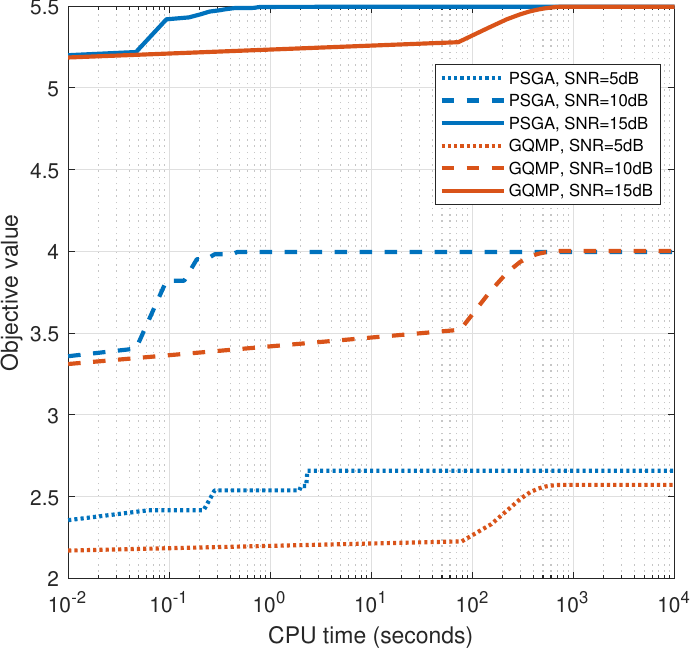}
    \caption{Convergence of PSGA and GQMP.}
    \label{SR_Convergence}
\end{figure}

We numerically evaluate the convergence of the proposed PSGA algorithm for solving the sum-rate maximization problem of SIC-free RSMA. Additionally, another algorithm that serves the same purpose using Generalized Quadratic Matrix Programming (GQMP) \cite{Jin_GQMP}, is incorporated for comparison. The simulation is under $N_T=2$, $K=2$ and QPSK as the constellation for all the streams. In Figure \ref{SR_Convergence}, it is evident that both algorithms eventually converge, but PSGA always converges with a much shorter computational time than GQMP, and sometimes converges at a higher objective value.

\begin{figure}
    \centering
    \includegraphics[width=7.2cm]{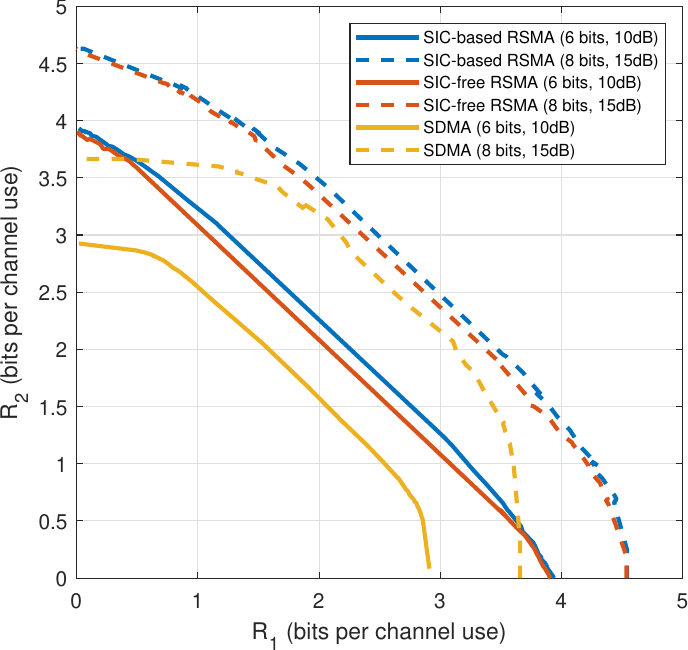}
    \caption{Rate region with $N_\text{T}=2$ and $K=2$, $\{\theta_1^\text{az}, \theta_2^\text{az}\} = \{0, \pi/9\}$, $\mathcal{K}=10\text{ dB}$.}
    \label{rate_region}
\end{figure}

By applying various weights to the users, the rate region can be obtained. Figure \ref{rate_region} depicts the ergodic rate region with $N_\text{T}=2$ and $K=2$ and indicates that RSMA achieves better rate regions than SDMA under finite constellations, similar to Gaussian distributed inputs \cite{Mao_EURASIP}. It also suggests that SIC-free RSMA performs similarly to SIC-based RSMA under different weights; hence, the two achieve similar rate regions. This result extends the conclusion in \cite{Sibo2024} from sum-rate to rate regions.

\begin{figure*}[!h]
    \centering
    \subfigure[$N_\text{T}=2$, $K=2$, $\{\theta_1^\text{az}, \theta_2^\text{az}\}= \{0,\pi/18\}$, $\kappa=10\text{ dB}$.]{\includegraphics[width=7.2cm]{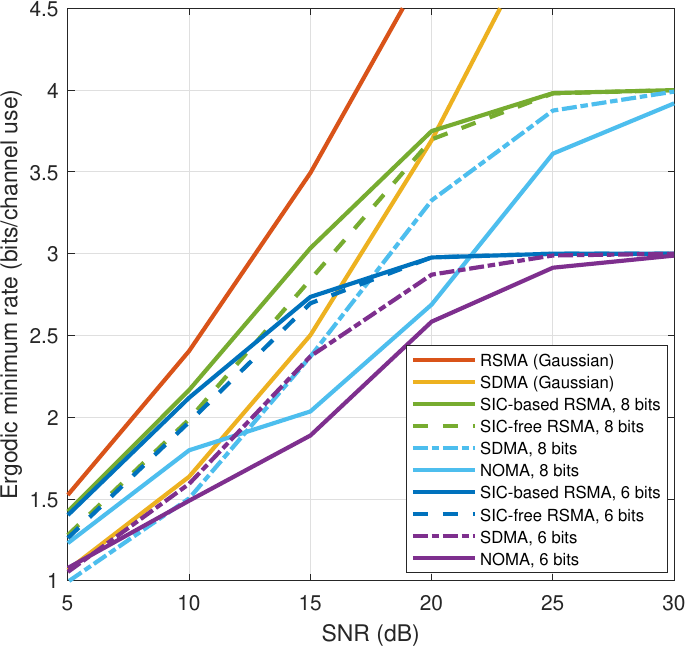}}\hspace{2cm}
    \subfigure[$N_\text{T}=4$, $K=3$, $\{\theta_1^\text{az}, \theta_2^\text{az}, \theta_3^\text{az}\}= \{-\pi/36,0,\pi/36\}$, $\kappa=10\text{ dB}$.]{\includegraphics[width=7.2cm]{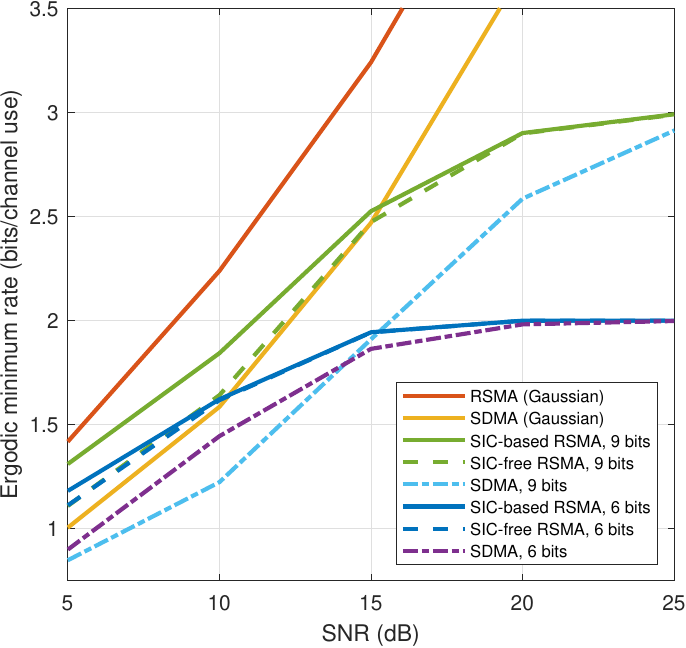}}
    \caption{Ergodic MMF performance.}
    \label{MM}
\end{figure*}

\subsection{Max-Min Fairness Performance}

Figure \ref{MM} depicts the MMF performance of SIC-based and SIC-free RSMA under finite constellations. The performance of SDMA under finite constellations is also depicted by fixing the transmission mode as in the WSR evaluations. For two users, we also included power-domain NOMA as a benchmark, where both users apply the same constellation as in SDMA, and the user ordering is determined by the channel strength\footnotemark[\value{footnote}].  In addition, the MMF performance of RSMA and SDMA under Gaussian distributed input signals, following the work in \cite{Mao_relay}, is included as a reference. It can be observed that RSMA schemes achieve better MMF performance than SDMA and NOMA under finite constellations; meanwhile, SIC-free RSMA preserves most of the MMF performance of SIC-based RSMA. This result extends the conclusion in \cite{Sibo2024} from sum-rate to MMF evaluations.

\subsection{Performance in Large-Scale Systems}
\begin{figure}[!h]
    \centering
    \includegraphics[width=7.2cm]{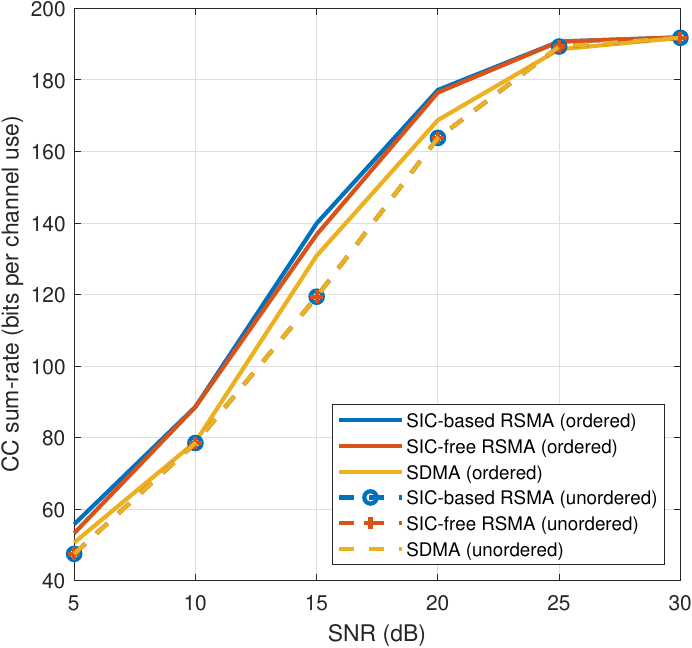}
    \caption{Ergodic sum-rate performance in large-scale systems with $N_\text{T}=128$, $K=64$, and $R_\text{max}=6$ bits per group.}
    \label{Large_SR_6bits}
\end{figure}
\begin{figure}[!h]
    \centering
    \includegraphics[width=7.2cm]{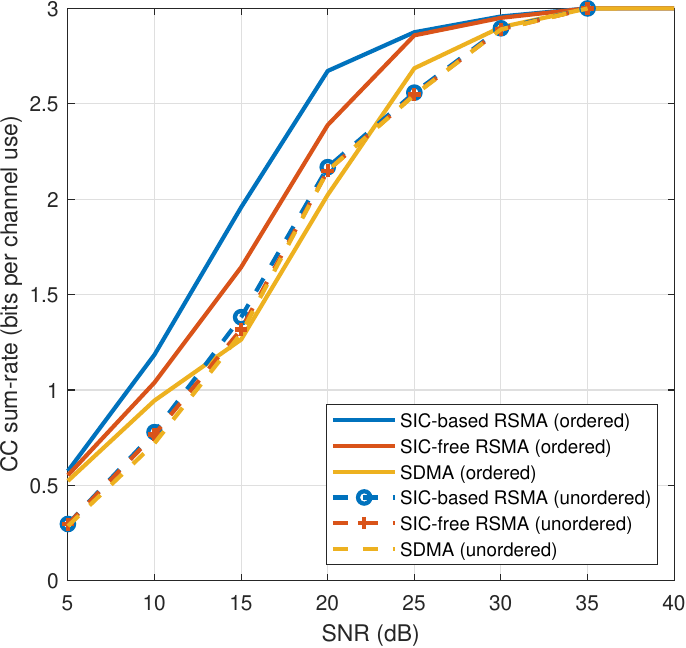}
    \caption{Ergodic MMF performance in large-scale systems with $N_\text{T}=128$, $K=64$, and $R_\text{max}=6$ bits per group.}
    \label{Large_MMF_6bits}
\end{figure}

We consider a large-scale rectangular antenna array with $N_\text{T}=128$, with 16 antennas in each row along the $y$-axis and 8 antennas in each column along the $z$-axis, serving 64 users. Figures \ref{Large_SR_6bits} and \ref{Large_MMF_6bits} depict the sum-rate and MMF performance of SDMA, SIC-based and SIC-free RSMA for large-scale systems, which is enabled by techniques including user grouping, null space projection, and dimensionality reduction introduced in Section \ref{Section_large_system}. We also evaluate the impact of user grouping by comparing the performance under the proposed grouping method in Algorithm \ref{Alg: User grouping} and random grouping, respectively labeled as "(ordered)" and "(unordered)" in the legends. For the channel model, $\theta_k^\text{az}$ and $\theta_k^\text{el}$, $\forall k \in \mathcal{K}$, are randomly drawn from the range of $[-\pi/4,\pi/4]$ and $[-\pi/6,\pi/6]$ with uniform distribution. $\kappa$ is set to 20dB to reflect the sparse nature of channels in typical large-scale antenna arrays. It can be observed that the conclusions drawn from small-scale systems as discussed previously, i.e., RSMA schemes outperform SDMA in both sum-rate and MMF, and SIC-free RSMA preserves most of the superiority of SIC-based RSMA, still holds in large-scale systems. Moreover, the proposed user grouping strategy generally enhances the performance of both SDMA and RSMA, with a more pronounced benefit for RSMA due to the high channel similarity within each user group.
\section{Conclusion}
In this work, we aimed to investigate the fundamental limits of SIC-based and SIC-free RSMA under finite constellations. We first derived the rate expressions for SIC-based and SIC-free RSMA. We then proposed precoding optimization algorithms for the WSR and MMF optimization problems. To enable these optimization algorithms in large-scale systems with feasible computational burden,  we introduced user grouping, null space projection and dimensionality reduction techniques. The numerical results show that RSMA schemes outperform SDMA under finite constellations and in both WSR or MMF evaluation, similar to the Gaussian distributed inputs. Furthermore, SIC-free RSMA preserves most of the superiority of SIC-based RSMA, which extends the conclusion in \cite{Sibo2024} from low-complexity precoders to fully optimized precoders, from sum-rate objective to WSR and MMF objectives, and from small-scale systems to large-scale systems.
\appendices
\section{Proof of Proposition \ref{WSR_prop}}\label{proof of WSR_prop}
If $\{u_1,...,u_K\}$ are all of the same value, it is trivial to see that $\mathbf{c}$ does not affect the objective function of $\mathcal{P}_1$. Hence, $C_i = R_\text{c}$ and $C_j = 0$, $\forall j \in \mathcal{K} \setminus i$, is one solution for $\mathbf{c}$ in $\mathcal{P}_1$.
If $\{u_1,...,u_K\}$ are of different values, we need the following lemma.

\begin{lemma}\label{WSR_lemma}
If $u_i > u_j$, $\forall j \in \mathcal{K}\setminus i$, for a given $\mathbf{P}$, one optimal $\mathbf{c}$ of $\mathcal{P}_1$ satisfies $C_i^\star = R_\text{c}$ and $C_j^\star = 0, \forall j \in \mathcal{K} \setminus \{i\}$.
\end{lemma}

\begin{IEEEproof}[Proof of Lemma \ref{WSR_lemma}]
Suppose $\mathbf{P}$ is given and the private rates, $R_{\text{p},k}, \forall k\in \mathcal{K}$, are not functions of $\mathbf{c}$, $\mathcal{P}_1$ is equivalent to
\begin{equation}
\begin{split}
    \mathcal{P}_{8}:\;\; \underset{\mathbf{c}}{\text{max}} \: \: & \sum_{k\in \mathcal{K}} u_k C_k\\ 
    \text{s.t.} \: \: \: \;
    & \sum_{k\in \mathcal{K}} C_k \leq R_\text{c}\\
    & \mathbf{c} \succcurlyeq \mathbf{0}^{1\times K},
\end{split}
\end{equation}
which is a simple linear programming and it is easy to see that Lemma \ref{WSR_lemma} is true.
\end{IEEEproof}

Suppose $(\mathbf{P}^\star, \mathbf{c}^\star)$ is optimal in $\mathcal{P}_1$, and $\mathbf{c}^\star$ does not satisfy $C_i = R_\text{c}$ and $C_j^\mathcal{K} = 0, \forall j \in \mathcal{K} \setminus i$. Then, according to Lemma \ref{WSR_lemma}, we can always find another $\mathbf{c}$ that gives a higher WSR than $\mathbf{c}^\star$ given that $\mathbf{P}=\mathbf{P}^\star$ in $\mathcal{P}_1$. By contradiction, $\mathbf{c}^\star$ must satisfy $C_i = R_\text{c}$ and $C_j = 0$, $\forall j \in \mathcal{K} \setminus i$. 

In the cases where $\arg \max_{k \in \mathcal{K}} u_k$ is a set of more than one element, a trivial extension of the above reasoning implies that any $\mathbf{c}$ allocating zero rate to users who are not most weighted is an optimal solution. This includes $C_i^\star = R_\text{c}$ and $C_j^\star = 0, \forall j \in \mathcal{K} \setminus i$, where $i \in \arg \max_{k \in \mathcal{K}} u_k$.

\section{Proof of Proposition \ref{c_opt_prop}}\label{proof of c_opt_prop}
$\mathcal{P}_4$ can be written as follows.
\begin{equation}
\begin{split}
    \mathcal{P}_{9}:\;\; \max_{\mathbf{c}}&\;\; \min_k C_k + R_{\text{p},k}\\
    \text{s.t.}
    &\;\; \sum_{k\in\mathcal{K}} C_k = R_\text{c}\\
    &\;\; \mathbf{c} \succcurlyeq \mathbf{0}^{1\times K}\\
\end{split}
\end{equation}
We introduce a slack variable, $\eta$, and convert $\mathcal{P}_{9}$ into
\begin{equation}
\begin{split}
    \mathcal{P}_{10}:\;\; \max_{\mathbf{c},t}&\;\;\;\;\;\; \eta\\
    \text{s.t.}
    &\;\; \sum_{k\in\mathcal{K}} C_k = R_\text{c}\\
    &\;\; \mathbf{c} \succcurlyeq \mathbf{0}^{1\times K}\\
    &\;\; C_k + R_{\text{p},k} \geq \eta, \forall k.
\end{split}
\end{equation}

Let $(\mathbf{c}^\star,\eta^\star)$ denote the solution of $\mathcal{P}_{10}$.
Without loss of generality, we assume $R_{\text{p},1}<R_{\text{p},2}<...<R_{\text{p},K}$, and the solution of $\mathcal{P}_{8}$ leads to $C_k^\star>0$, $\forall k \in \{1,...,k'\}$ with $1\leq k'\leq K$, and $C_k^\star=0$, $\forall k\in\{k'+1,...,K\}$. We have the following lemma.
\begin{lemma}\label{common_allocation_lemma}
    $C_k^\star + R_{\text{p},k} = \eta^\star$, $\forall k\in\{1,...,k'\}$, and with some value for $\eta^\star$.
\end{lemma}
\begin{IEEEproof}
    It is trivial to see that $R_{\text{p},k}<\eta^\star$, $\forall k\in\{1,...,k'\}$. Let $\mathcal{K}_1$ and $\mathcal{K}_2$ be a partition of $\{1,...,k'\}$. Suppose $(\mathbf{c}^\star,\eta^\star)$ leads to $C^\star_k + R_{\text{p},k} > \eta^\star$ for $k\in\mathcal{K}_1$, and $C^\star_k + R_{\text{p},k} = \eta^\star$ for $k\in\mathcal{K}_2$. Then, another choice of $\mathbf{c}$, denoted by $\mathbf{c}^{\diamond}$, can be found by $\mathbf{c}^\diamond=\alpha\mathbf{c^\circ}$, where $C^\circ_k=(C^\star_k - R_{\text{p},k} + \eta^\star)/2$ for $k\in\mathcal{K}_1$, $C^\circ_k=C_k^\star$ for $k\in\mathcal{K}_2$, and $\alpha = R_\text{c}/ \sum_{k\in\mathcal{K}}C^\circ_k$. It is trivial to see that $\alpha>1$. Therefore, $\min_{k\in\{1,...,k'\}} C^\circ_k + R_{\text{p},k} > \min_{k\in\{1,...,k'\}} C^\star_k + R_{\text{p},k}$, which indicates that $\mathbf{c}^\star$ is not the solution of $\mathcal{P}_{10}$. By contradiction, Lemma \ref{common_allocation_lemma} is proved.
\end{IEEEproof}
    
Lemma \ref{common_allocation_lemma} can be re-written as
\begin{equation}\label{linear equation}
    \mathbf{c}' + \mathbf{d} = \eta\mathbf{1}^{1\times k'},
\end{equation}
where $\mathbf{d}=[R_{\text{p},1},...,R_{\text{p},k'}]$, and $\mathbf{c}'=[C_1,...,C_{k'}]$. Combining (\ref{linear equation}) and the constraint on $\sum_{k\in\mathcal{K}} C_k = R_\text{c}$, the global optimal solution of $\mathbf{c}'$ and $\eta$, denoted by $\mathbf{c}'^\star$ and $\eta^\star$, can be then obtained by
\begin{equation}\label{c_opt}
    \mathbf{c}'^\star = \eta^\star\mathbf{1}^{1\times k'} -\mathbf{d},
\end{equation}
and
\begin{equation}\label{t_opt}
    \eta^\star = \frac{R_\text{c}+\mathbf{d}\mathbf{1}^{k'\times1}}{k'}.
\end{equation}

Although $k'$, i.e., the number of users whose are strictly served by the common stream, is still not known, we can test it using (\ref{c_opt}) and (\ref{t_opt}). This is because allocating the common stream rate based on (\ref{c_opt}) and (\ref{t_opt}) to more users than $k'$ will lead to $\mathbf{c}'$ containing negative entries, which is infeasible. Hence, by iteratively guessing the number of users to be served by the common stream, Algorithm \ref{c_opt alg} is obtained.

\section{Transmission Mode Dictionaries Used For Numerical Results}\label{Appendix_Tx_mode}
The transmission mode dictionaries used to produce the numerical results in Section \ref{Section_results} are depicted in Tables \ref{Table: Tx mode 2 UE 6 bits}-\ref{Table: Tx mode 3 UE 9 bits}.
\begin{table}[!h]
    \centering
    \caption{Transmission mode dictionary for $K=2$ and $R_\text{max}=6$ bits.}
    \begin{tabular}{|c|c|c|c|c|c|}
        \hline
        Mode & 1 & 2 & 3 & 4\\
        \hline
        $s_\text{c}$ & - & QPSK & 16QAM & 64QAM\\
        \hline
        $s_{\text{p},k}$ & 8QAM & QPSK & BPSK& -\\
        \hline
    \end{tabular}
    \label{Table: Tx mode 2 UE 6 bits}
    \vspace{0.6cm}
    
    \centering
    \caption{Transmission mode dictionary for $K=2$ and $R_\text{max}=8$ bits.}
    \begin{tabular}{|c|c|c|c|c|c|}
        \hline
        Mode & 1 & 2 & 3 & 4 & 5\\
        \hline
        $s_\text{c}$ & - & QPSK & 16QAM & 64QAM & 256QAM\\
        \hline
        $s_{\text{p},k}$ & 16QAM & 8QAM & QPSK & BPSK& -\\
        \hline
    \end{tabular}
    \label{Table: Tx mode 2 UE 8 bits}
    \vspace{0.6cm}

    \centering
    \caption{Transmission modes dictionary for $K=3$ and $R_\text{max}=6$ bits.}
    \begin{tabular}{|c|c|c|c|c|c|}
        \hline
        Mode & 1 & 2 & 3\\
        \hline
        $s_\text{c}$ & - & 8QAM & 64QAM\\
        \hline
        $s_{\text{p},k}$ & QPSK & BPSK& -\\
        \hline
    \end{tabular}
    \label{Table: Tx mode 3 UE 6 bits}
    \vspace{0.6cm}

    \centering
    \caption{Transmission modes dictionary for $K=3$ and $R_\text{max}=9$ bits.}
    \begin{tabular}{|c|c|c|c|c|c|}
        \hline
        Mode & 1 & 2 & 3 & 4\\
        \hline
        $s_\text{c}$ & - & 8QAM & 64QAM & 512QAM\\
        \hline
        $s_{\text{p},k}$ & 8QAM & QPSK & BPSK& -\\
        \hline
    \end{tabular}
    \label{Table: Tx mode 3 UE 9 bits}
    \vspace{0.6cm}

\end{table}

\bibliographystyle{IEEEtran}
\bibliography{references}

\end{document}